\documentclass[aps, 11pt, onecolumn,
floatfix, fleqn, notitlepage, nofootinbib,
amsmath, amssymb, amsfonts,
preprintnumbers, showpacs, showkeys,longbibliography]{revtex4-1}

\usepackage[colorlinks=true, linkcolor=blue, citecolor=blue,
            filecolor=blue, urlcolor=blue]{hyperref}
\usepackage{dcolumn, graphicx, moreverb}
\newcolumntype{.}{D{.}{.}{-1}} 
\usepackage[american]{babel}

\usepackage{mathptmx}
\usepackage[T1]{fontenc}
\usepackage{multirow}

\DeclareFontFamily{U}{wncy}{}
\DeclareFontShape{U}{wncy}{m}{n}{<->wncyr10}{}
\DeclareSymbolFont{mcy}{U}{wncy}{m}{n}
\DeclareMathSymbol{\Sh}{\mathord}{mcy}{"58} 

\newcommand{\q}[2]{\ensuremath{#1\ \mathrm{#2}}} 

\begin{document}

\title{Effect of pulsed hollow electron-lens operation on the proton beam core in LHC}

\author{M.~Fitterer}
\email[Email:]{mfittere@fnal.gov}
\author{G.~Stancari}
\author{A.~Valishev}

\affiliation{Fermi National Accelerator Laboratory, PO Box 500,
  Batavia, Illinois 60510, USA}
\thanks{Fermilab is operated by Fermi Research Alliance, LLC under
	Contract No.~DE-AC02-07CH11359 with the United States Department of
	Energy. This work was partially supported by the US DOE LHC
	Accelerator Research Program (LARP) and by the European FP7 HiLumi
	LHC Design Study, Grant Agreement 284404.}
\date{Nov.~7, 2016}

\begin{abstract}
Collimation with hollow electron beams is currently one of the most promising concepts for active halo control in the HL-LHC. In order to further increase the diffusion rates for a fast halo removal as e.g. desired before the squeeze, the electron lens (e-lens) can be operated in pulsed mode. In case of profile imperfections in the electron beam the pulsing of the e-lens induces noise on the proton beam which can, depending on the frequency content and strength, lead to emittance growth. In order to study the sensitivity to the pulsing pattern and the amplitude, a beam study (machine development MD) at the LHC has been proposed for August 2016 and we present in this note the preparatory simulations and estimates.
\end{abstract}

\preprint{FERMILAB-TM-2635-AD}
\maketitle

\clearpage
\tableofcontents
\clearpage

\section{Introduction}

For high energy and high intensity hadron colliders like the HL-LHC, halo control becomes more and more relevant if not necessary for a safe machine operation and control of the targeted stored beam energy, which lies in the range of several hundred MJ in case of the HL-LHC. Past experiments at the Fermilab Tevatron proton-antiproton collider demonstrated a successful halo control with hollow electron beams together with a high reliability of the device itself \cite{giuliotevatron}, and the hollow electron lens (HEL) is currently considered the best suited device for an active halo control \cite{roderikelensreview}.

Based on operational experience and study of failure scenarios for HL-LHC, the main concerns for HL-LHC for which the HEL would represent a mitigation measure are \cite{elensreview}:
\begin{itemize}
	\item Loss spikes have been observed in 2012 during the squeeze and adjust. Cleaning the tails with a HEL would mitigate these loss spikes and thus improve machine availability.
	\item The LHC tails are overpopulated compared to a Gaussian beam profile \cite{gianlucaelensreview}. Scaling to HL-LHC bunch intensities and energy is not obvious. The HEL would offer better control of the energy stored in the tails and thus more operational margin.
	\item Orbit distortions arising from earthquakes or the Geothermie2020 project, a project to explore geothermal energy in the Geneva region \cite{vibrationhalfday,jorgchamonix,michaelaelensreview}.
	\item Crab cavity failures induce large orbit distortions and a depletion of the tails with the HEL would thus be necessary for machine protection \cite{ramaelensreview,danielelensreview}.
\end{itemize}
This implies that a fast halo removal after the ramp is needed in order to deplete efficiently the halo at flat-top before the squeeze/adjust, and a continuous halo removal during stable beams in order to control the halo sufficiently and mitigate losses due to e.g. orbit distortions and provide more margin for fast losses from e.g. crab cavity failures.

To estimate the halo removal rates for the different HL-LHC scenarios, first numerical simulations for the nominal LHC \cite{elenscdr,valentinaelenslhc,valishevelenslhc} and the HL-LHC \cite{fittererhalo} have been conducted. The simulations for HL-LHC show halo removal rates \footnote{The halo removal rate is here defined as the relative intensity loss for a uniform transverse distribution between 4 and $6~\sigma$ and Gaussian distribution in $z,\frac{\Delta p}{p_0}$.} from $3~\%/$min to $9~\%/$min at flat-top and without collisions, and from $12~\%/$min to $18~\%/$min during $\beta^*$-leveling and with collisions. For a continuous halo removal during $\beta^*$-leveling the halo removal rates are sufficient, while for a fast halo removal at flat-top the removal rates are too small for a fast and efficient removal within 10-20 minutes. The halo removal rates can in general be increased by pulsing the e-lens \cite{tevatronabortgap,valishevelenslhc}. Two different pulsing patterns are currently considered:
\begin{itemize}
	\item \textbf{random:} the electron beam current is modulated randomly,
	\item \textbf{resonant:} the e-lens is switched on only every $n$th turn with $n=2,3,4,\ldots$.
\end{itemize}
One of the main reservations about pulsing the e-lens is the possibility of emittance growth due to noise induced on the beam core by the e-lens. In this note we will only consider the contribution from dipole kicks. Higher order kicks are present, but their effect is estimated to be much smaller. For an ideal radially symmetric hollow electron lens with an S-shaped geometry, the beam core would not experience any dipole kick. This is because, the kicks due to the bends from the e-lens compensate each other in case of an S-shaped geometry and for radially symmetric e-lens profiles the field in the center of the hollow e-lens is zero \footnote{Note that for higher orders the kicks are not compensated, also in case of an S-shaped geometry.}. The amplitude of the dipolar noise is therefore given by the electron beam profile imperfections, for which an estimate is given in Sec.~\ref{sec:noiseamp}.

Emittance growth due to noise in general depends on:
\begin{itemize}
	\item noise amplitude
	\item noise spectrum
	\item machine non-linearities and beam configuration (separated/colliding beams)
	\item transverse feedback system
\end{itemize}

In case of a random pulsing pattern, the effect can be estimated with the well known theoretical formulas and simulation models for emittance growth due to white noise \cite{Lebedevnoisessc,alexahinnoise,ohminoise}, which have been compared also to experimental results at the LHC \cite{bbnoiseLHC}. The simulations and theoretical formulas seem to reproduce the experimental results well within a factor two \cite{bbnoiseLHC}. The source of the factor two is currently unknown. One possibility could be noise induced by the feedback system itself, which has not been taken into account in the simulations and theoretical estimates.

For the resonant pulsing patterns on the other hand no theoretical estimates exist as the emittance growth is determined by the non-linearities present. Pulsing the e-lens every $k$th turn and considering only the dipole kick, will drive all harmonics of the $k$th order resonance. This can be seen by writing down the Fourier series of an excitation every $k$th turn~\footnote{ 
The Dirac comb is constructed from Dirac delta functions
\begin{eqnarray}	
\Sh_T(t)&=&\sum_{p=-\infty}^{+\infty}\delta(t-pT),
\end{eqnarray}
and its Fourier series is given by:
\begin{eqnarray}	
\Sh_T(t)&=&\frac{1}{T}\sum_{n=-\infty}^{+\infty}e^{\frac{2\pi int}{T}}.
\end{eqnarray}
Using the expression for the Fourier series of a Dirac comb and the relation
\begin{equation}
	\Sh_{kT}(t)=\frac{1}{k}\Sh_T(\frac{t}{k}),
\end{equation}
the Fourier series for an excitation every $k$th turn can be derived.}, where the excitation can be represented by:
\begin{eqnarray}\label{intro:eqn:1}
f(t)&=&\sum_{p=-\infty}^{+\infty}\delta(t-p(kT)),
\end{eqnarray}
and its Fourier series by:
\begin{eqnarray}\label{intro:eqn:2}
f(t)=\Sh_{kT}(t)
&=&\frac{1}{kT}\sum_{n=-\infty}^{+\infty}e^{2\pi i f_nt} \ \text{with} \ f_n=\frac{n}{k}f_{\rm rev}.
\end{eqnarray}
Note that each harmonic has the same amplitude, explicitly $\frac{1}{kT}$ and that the amplitude for the different pulsing patterns decreases like $\frac{1}{k}$.

To identify the limits on the noise amplitude and the pulsing pattern specifically for the LHC and HL-LHC a MD was proposed for August 2016. The MD setup is described in detail in Sec.~\ref{sec:md}. In preparation of the MD, the different scenarios have been simulated with the tracking code Lifetrac \cite{lifetrac}. The goal of these simulations is to identify pulsing patterns that are most dangerous in terms of emittance growth, and pulsing patterns that are most efficient for halo control. The expected noise amplitudes from the HEL are summarized in Sec.~\ref{sec:noiseamp} and the limits on the noise amplitude obtained from simulations are described in Sec.~\ref{sec:sim}. As the emittance growth depends on the machine non-linearities present, the MD represents also a good check of the machine model used for the simulations and the prediction accuracy of this model. For the effects of pulsing on the halo removal rates, see Ref.~\cite{fittererhalo}.

\section{Overview of the experimental proposal}
\label{sec:md}
In order to keep the machine changes minimal and to also be able to quickly refill the machine in case of beam loss, the MD is conducted with 48 single bunches, single beam and at standard injection settings:
\begin{itemize}
	\item injection energy (450 GeV)
	\item single bunch intensity: $0.7\times10^{11}$, number of bunches: $48$
	\item normalized emittance: $2.5~\mathrm{\mu m}$, bunch length ($4~\sigma$): 1.0~ns
	\item injection optics ($\beta^*=11$~m), injection tunes
	\item chromaticity: $Q'_{x/y}=+15$ (standard 2016 settings)
	\item Landau damping octupole current of $I_{\mathrm{MO}}=\pm19.6~\mathrm{A}$, explicitly +19.6 A for MOF circuit and -19.6~A for MOD circuit (standard 2016 settings)	
\end{itemize}
In order to minimize the emittance blow-up due to intra-beam scattering, a smaller bunch intensity of $0.7\times10^{11}$ is requested for the MD. The lower limit of $0.7\times10^{11}$ is in this case determined by the orbit correction system, for which the BPMs only deliver a good signal for bunch intensities above $0.5\times10^{11}$. In order to be also more sensitive to the relative emittance growth, the HL-LHC normalized emittance of $2.5~\mathrm{\mu m}$ is requested which is smaller than the nominal LHC single bunch emittance.

\begin{figure}[h]
	\centering
	\includegraphics[width=0.9\linewidth]{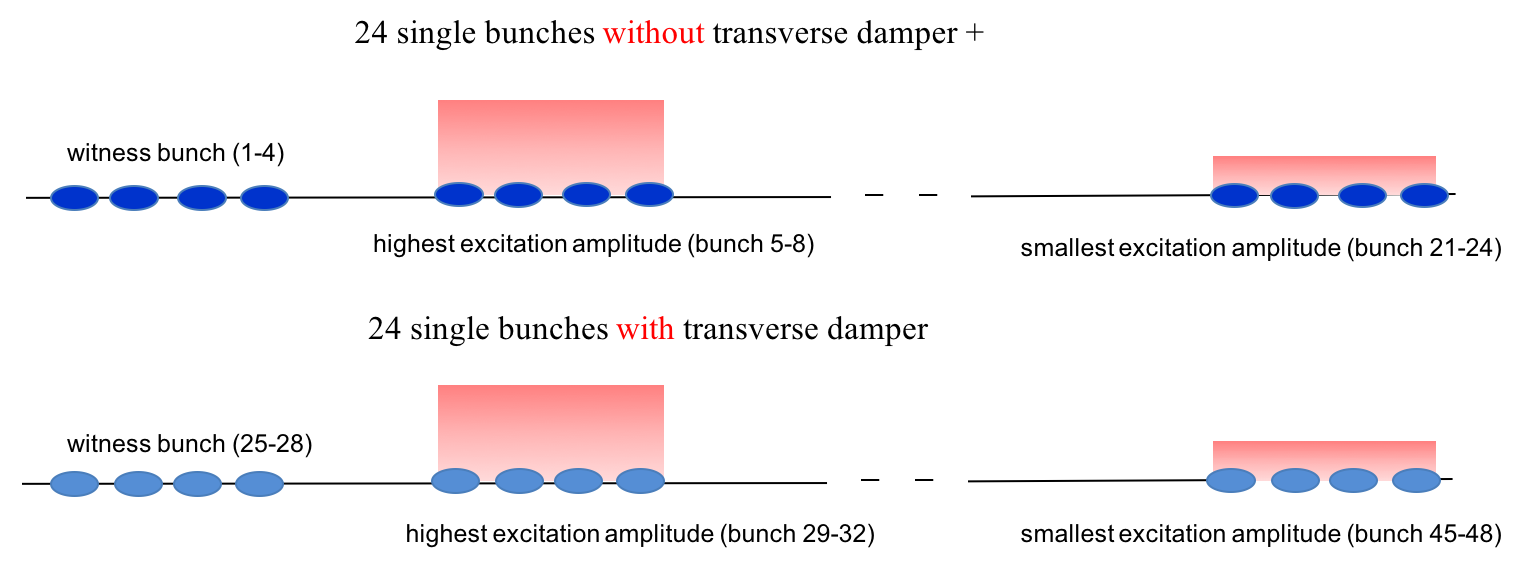}
	\caption{Proposed filling scheme for the MD. For the dark blue bunches (first 24) the transverse damper is active, for the light blue bunches (second 24) the transverse damper is not active. The excitation amplitude indicated with a red rectangle is constant over each group of 4 bunches in order to later be able to increase the statistics by averaging over several bunches. With this filling scheme in total 5 different excitation amplitudes plus the case of no excitation and the cases with the transverse active and not active can be studied during the same fill.}
	\label{md:fig:1}
\end{figure}

The noise induced by a pulsed e-lens can be to first order approximated by a dipole kick with the corresponding noise pattern/frequency spectrum (see Sec.~\ref{sec:noiseamp}). In case of the LHC almost arbitrary noise spectra seen by the \emph{whole beam} can be generated using the transverse damper (ADT) and the amplitude of the noise seen by \emph{individual} bunches can be controlled by a windowing function placed on top of the generated excitation. This implies that, for each fill, only one excitation pattern with varying amplitude can be studied. In order to have full flexibility in the control of the amplitude\footnote{The minimum rise time of the ADT kicker is 700~ns, which determines the minimum bunch spacing required to control the noise amplitude of each individual bunch. By injecting individual bunches the bunch spacing can be chosen between 250~ns to 1~$\mu$s \cite{giulia}.} and to avoid multi-bunch instabilities, the MD is conducted with single bunches. Based on these considerations the filling scheme illustrated in Fig.~\ref{md:fig:1} has been chosen for the MD. The filling scheme comprises $2\times 4$ witness bunches (4 with and 4 without transverse damper) and $2\times 2$ bunches per amplitude (2 with and 2 without damper). As each group of 4~bunches experiences the same or no excitation, the statistical significance of the results can be improved by averaging over each group.

In total, three different pulsing patterns are intended to be studied in the MD: the two pulsing patterns exhibiting the largest emittance growth - pulsing every 7th and 10th turn - and one pulsing pattern showing no emittance growth, e.g. pulsing every 8th turn (see Sec.~\ref{sec:sim}).

\section{Sources of noise and estimate of noise amplitude}
\label{sec:noiseamp}
If the e-lens is operated in pulsed mode, noise can be induced on the circulating proton beam by uncompensated kicks at small amplitudes. With the current e-lens layout \cite{elenscdr}, parasitic kicks on the core can arise due to profile imperfections in the electron beam in the central region (main solenoid) and at the entrance and exit of the e-lens. As derived in detail in Sec.~\ref{noiseamp:sec:1} and Sec.~\ref{noiseamp:sec:2} the contribution from the central region is dominating. The total estimated integrated kick on axis ($x_p=y_p=x_p'=y_p'=0$) from profile imperfections and the e-lens bends assuming the current e-lens design parameters \cite{elenscdr} of $B_{\mathrm{solenoid}}=\q{5}{T}$, $I_e=\q{5.0}{A}$, $E_{e} = \q{10.0}{keV}$, $l_{\rm e-lens}=\q{3}{m}$ and $E_{p} = \q{7.0}{TeV}$ is:
\begin{equation}
\Delta x', \Delta y'= \q{15}{nrad}
\end{equation}
where $B_{\mathrm{solenoid}}$ is the magnetic field strength of the main solenoid, $E_{\rm e}$ the electron beam energy, $E_{\rm p}$ the proton beam energy and $l_{\rm e-lens}$ the length of the e-lens.

\subsection{Noise due to uncompensated kicks from e-lens bends}
\label{noiseamp:sec:1}
The symplectic map for an e-lens bend is derived in Ref.~\cite{giulioelensbends}. In this case the e-lens bends are modeled as a cylindrical pipe with a static charge distribution of 1~A, 5~keV electrons. In this model the magnetic field and the compression of the electron beam density by the solenoid field are neglected. Neglecting the electron beam velocity leads to an underestimation of the kick by a factor $\beta_e\beta_p= 0.2$ for 10~keV electrons\footnote{Note that $F_{\mathrm{Lorentz}}\sim (1+\beta_e\beta_p)\cdot F_{\mathrm{electric}}$}, while the missing compression leads to an overestimation of the kick as the increase of the electron beam size towards the start/end of the electron lens is not considered. For an S-shaped e-lens, the transverse dipole kicks at entrance and exit compensate each other to first order. Uncompensated kicks therefore arise due to profile imperfections. As a first estimate, we assume 10\% fluctuations between the entrance and exit, which can originate from profile imperfections, current fluctuations etc., and that the kicks from entrance and exit due to this fluctuation add up. Neglecting magnetic effects the kick is given by:
\begin{equation}\label{noiseamp:eqn:1}
\Delta p_{x,y}=\frac{q}{v_z}\int_{z_1}^{z_2}E_{x,y}dz \quad \Rightarrow \quad \Delta x',\Delta y' = \frac{1}{(B\rho)_p\cdot v_z}\int_{z_1}^{z_2}E_{x,y}dz,
\end{equation}
where $q,v$ and $(B\rho)_p$ are the charge, velocity and magnetic rigidity of the circulating proton beam and $z=v_z\cdot t$ the longitudinal position. For a 1~A, 5~keV electron beam and 7 TeV proton beam, the integrated electric field and kick is given by
\begin{equation}
\int_{z_1}^{z_2} E_{x} dz= 10 \ \mathrm{kV} \Rightarrow \Delta x'= \q{1.4}{nrad}.
\end{equation}
This corresponds to
	\footnote{The electron beam charge distribution $\rho_e$, line density $\lambda_e$, current $I_e$ and energy $E_{\mathrm{kin},e}$ are related by:
	\begin{equation}
	\begin{array}{l}
	I_e=\lambda_e\cdot \beta_e \cdot c \ \text{and} \ \lambda_e=\rho_e\cdot A \
	\mathrm{ with } \  \beta_e=\sqrt{1-\frac{m_ec^2}{(E_{\mathrm{kin},e}+m_ec^2)}}\ \mathrm{ and } \ A = \pi(r_{a}^2-r_{i}^2) = \mathrm{const.}
	\end{array}
	\end{equation}
	where $A$ is the area of the electron beam (here an annular uniform profile), $m_e$ the rest mass of the electron and $c$ the speed of light. Under these assumption the beam charge distribution scales with the energy and beam current as:
	\begin{equation}
	\rho_e(E_{\mathrm{kin},e,2},I_{e,2})=\frac{I_{e,2}\beta_{e,1}}{I_{e,1}\beta_{e,2}}\rho_e(E_{\mathrm{kin},e,1},I_1).
	\end{equation}
	As the electron beam distribution is assumed to be static in this model, the electric field is proportional to the charge density and therefore obeys the same scaling rule:
	\begin{equation}\label{eqn:ecurrenscale}
	E_{x/y}(E_{\mathrm{kin},e,2},I_{e,2})=\frac{I_{e,2}\beta_{e,1}}{I_{e,1}\beta_{e,2}}E_{x/y}(E_{\mathrm{kin},e,1},I_1).
	\end{equation}
	}
\begin{equation}
\int_{z_1}^{z_2} E_{x} dz= 36 \ \mathrm{kV} \Rightarrow \Delta x'= \q{5.1}{nrad}
\end{equation}
for a 5~A, 10~keV electron beam (the e-lens design parameters \cite{elenscdr}). \pagebreak Note that the vertical electric field vanishes in the ideal case. As the electric field scales linearly with the electron beam current, the uncompensated kick for either entrance or exit in the horizontal and vertical plane assuming 10\% fluctuation yields
\begin{equation}
\Delta x', \Delta y'= \q{0.5}{nrad}.
\end{equation}

\subsection{Central region (main solenoid)}
\label{noiseamp:sec:2}
\begin{figure}[h]
	\centering
	\includegraphics[width=0.5\linewidth]{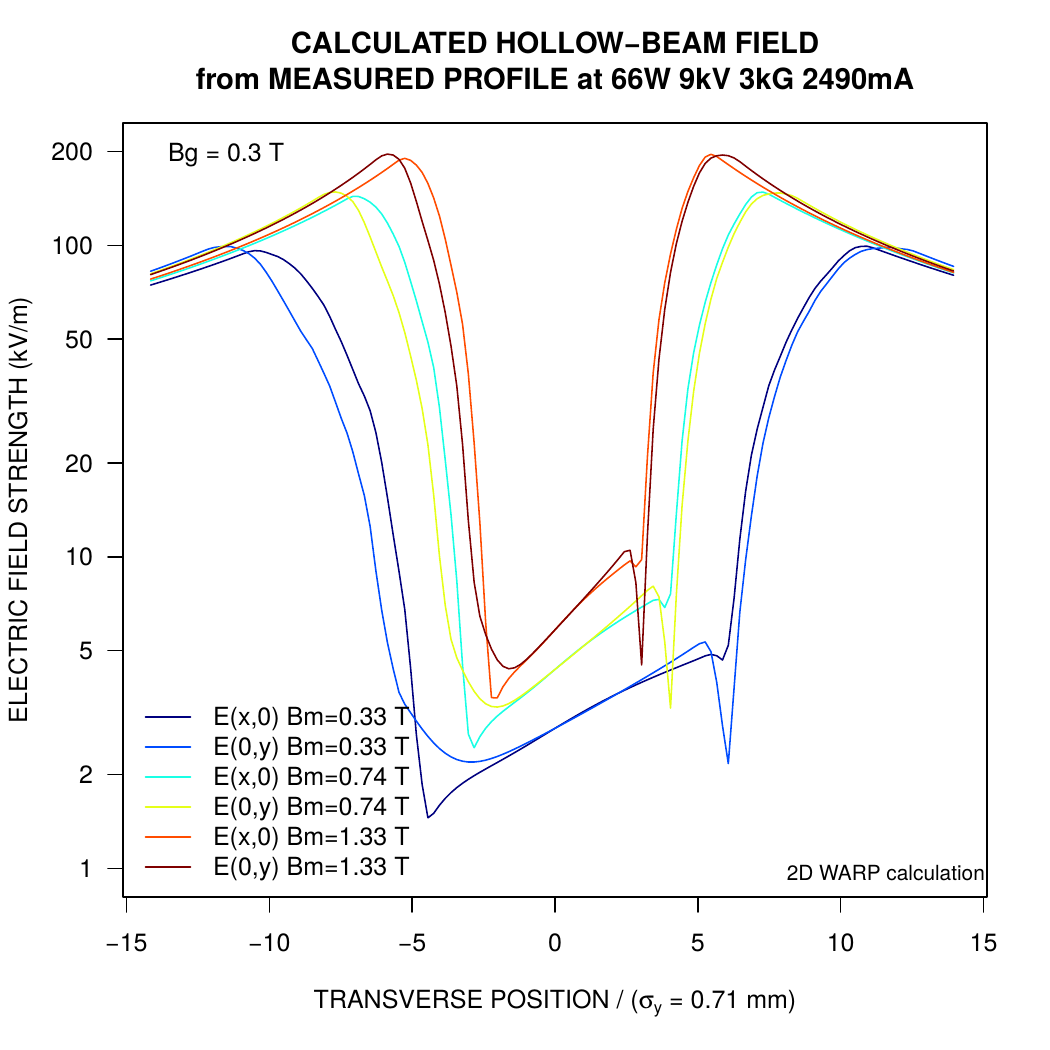}
	\caption{Calculated hollow electron beam field from measured profile of 9~kV, 2.49~A e-gun and different main solenoidal fields $Bm$. The field has been calculated using the code WARP \cite{warp}.}
	\label{noiseamp:fig:1}
\end{figure}

For a perfectly uniform, annular and radially symmetric profile, the field for $r<R_1$ vanishes, where $R_1$ is the inner radius of the hollow electron beam and $r$ is the radial amplitude of the proton beam particle. In case of electron beam profile imperfections the radial symmetry is broken, leading to a residual field at the beam core. Fig.~\ref{noiseamp:fig:1} shows an example of the electric field calculated from a measured asymmetric profile.

A first estimate of the residual kick for the e-lens design parameters with a main solenoid field of $B_{\mathrm{solenoid}}=\q{5}{T}$, $I_e=\q{5.0}{A}$, $E_{\rm e} = \q{10.0}{keV}$, $l_{\rm e-lens}=\q{3}{m}$ and $E_{\rm p} = \q{7.0}{TeV}$ \cite{elenscdr} can be obtained by scaling the electric field at the center illustrated Fig.~\ref{noiseamp:fig:1}. We will first derive the scaling with the solenoid field $B_{\mathrm{solenoid}}$ in Sec.~\ref{noiseamp:sec:2:1} and then the electron beam current $I_e$ and energy $E_{e}$ in Sec.~\ref{noiseamp:sec:2:2}. Summarizing both scalings, the integrated kick at the center of the proton beam is estimated to be:
\begin{equation}
\Delta x', \Delta y'(10~\mathrm{keV},5~\mathrm{T},2.5~\mathrm{A})=\q{15}{nrad}
\end{equation}

\subsubsection{Scaling with the main solenoid field}
\label{noiseamp:sec:2:1}
\begin{figure}[h]
	\centering
	\includegraphics[width=0.5\linewidth]{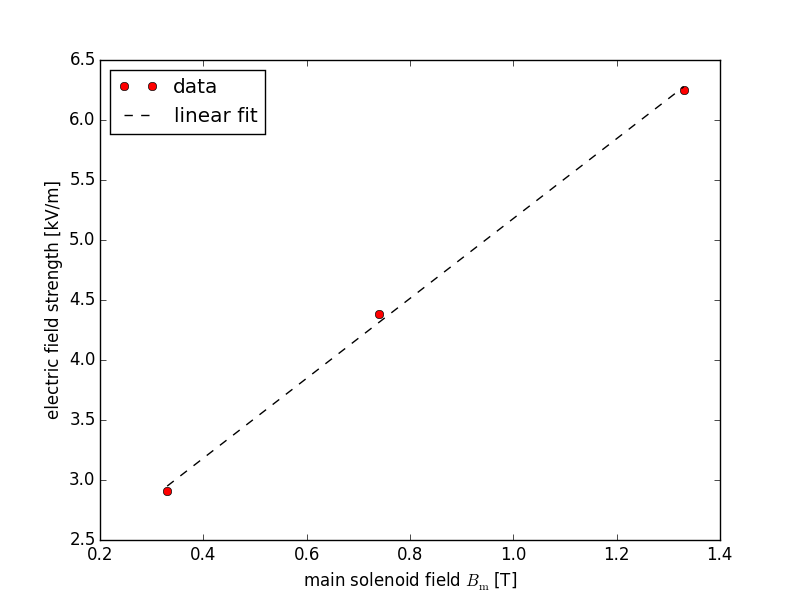}
	\caption{Calculated hollow electron beam electric field from measured profile of a 9~kV, 2.49~A e-gun and different main solenoid fields $Bm$. The field has been calculated using the code WARP \cite{warp}.}
	\label{noiseamp:fig:2}
\end{figure}

Due to the magnetic compression the density of the electron beam increases and therefore the electric field on axis $E_{\mathrm{center}}$ grows linearly with the magnetic field of the main solenoid $B_{\mathrm{solenoid}}$
	\footnote{
	We want to estimate how the electric field due to the magnetized electron beam changes in response to a change in the solenoid field. We start by considering two infinitely long axially symmetric systems, one where the solenoid field is $B_{1,\mathrm{solenoid}}$ and the other where it is scaled by a factor $k$:
	\begin{equation}
	B_{2,\mathrm{solenoid}}=k\cdot B_{1,\mathrm{solenoid}}
	\end{equation}
	Because the electron beams are magnetized, their radii are related by:
	\begin{equation}
	B_{2,\mathrm{solenoid}}\cdot r_2^2=B_{1,\mathrm{solenoid}}\cdot r_1^2 \ \Rightarrow \ r_2=\frac{1}{\sqrt{k}}\cdot r_1
	\end{equation}
	We impose scaling of all coordinates, including the longitudinal:
	\begin{equation}
	\mathbf{r}_2=\frac{1}{\sqrt{k}}\mathbf{r}_1
	\end{equation}
	For a fixed number of particles each of a given charge, the potential is:
	\begin{equation}
	\Phi(\mathbf{r})=\frac{1}{4\pi\epsilon_0}\sum_i\frac{q_i}{\left|\mathbf{r}-\mathbf{r_i}\right|}
	\end{equation}
	This leads to the following scaling of the potential:
	\begin{eqnarray}\label{eqn:phiscale}
	\Phi(\mathbf{r}_2)&=&\frac{1}{4\pi\epsilon_0}\sum_i\frac{q_i}{\left|\mathbf{r}_2-\mathbf{r}_{2,i}\right|}=\frac{1}{4\pi\epsilon_0}\sum_i\frac{q_i}{\frac{1}{\sqrt{k}}\left|\mathbf{r}_1-\mathbf{r}_{1,i}\right|}\\
	&=&\sqrt{k}\cdot \Phi(\mathbf{r}_1)
	\end{eqnarray}
	The whole system is compressed by $\frac{1}{\sqrt{k}}$, including the coordinates of the charges $\mathbf{r}_{i}$, explicitly $\mathbf{r}_{2,i}=\frac{1}{\sqrt{k}}\mathbf{r}_{1,i}$. The electric field can then be derived from the electric potential $\Phi(\mathbf{r})$ by:
	\begin{equation}
	\mathbf{E}(\mathbf{r})=-\nabla_{\mathbf{r}} \Phi(\mathbf{r})
	\end{equation}
	yielding:
	\begin{eqnarray}
	\mathbf{E}(\mathbf{r_2})&=&-\nabla_{\mathbf{r_2}} \Phi(\mathbf{r_2}) \overset{\nabla_{\mathbf{r_2}}=\sqrt{k}\cdot\nabla_{\mathbf{r_1}},\mathrm{Eqn.}~\ref{eqn:phiscale}}{=} \sqrt{k}\cdot\nabla_{\mathbf{r_1}}\sqrt{k}\cdot\Phi(\mathbf{r}_1)\\
	&=& k\cdot\nabla_{\mathbf{r_1}}\Phi(\mathbf{r}_1) = k\cdot \mathbf{E}(\mathbf{r_1}).
	\end{eqnarray}
	The electric field seen by the proton beam thus scales as the main solenoid field. This scaling will break down at low magnetic fields because the beam will cease to be fully magnetized.
	}
:
\begin{equation}
E_{\mathrm{center}} = a + m\cdot B_{\mathrm{solenoid}}
\end{equation}
This is illustrated in Fig.~\ref{noiseamp:fig:2} using the values of the measured profile for the different values of the main solenoid magnetic field $B_{\mathrm{solenoid}}$ shown in Fig.~\ref{noiseamp:fig:1}. The linear fit of  $B_{\mathrm{solenoid}}$ versus $E_{\mathrm{center}}$ yields:
\begin{equation}
a = 3.33 \ \mathrm{kV/m}, \ m=1.85 \ \mathrm{kV/Tm}
\end{equation}
leading to a field of $E_{\mathrm{center}}(9~\mathrm{keV},5~\mathrm{T},2.5~\mathrm{A}) = 18.5~\mathrm{kV/m}$ for $B_{\mathrm{solenoid}}=5$~T. The integrated electric field $E_{\mathrm{center, integrated}}$ for an e-lens of 3~m is then:
\begin{equation}
 E_{\mathrm{center, integrated}}(9~\mathrm{keV},5~\mathrm{T},2.5~\mathrm{A}) = 55.5~\mathrm{kV}.
\end{equation}

\subsubsection{Scaling with the electron beam current}
\label{noiseamp:sec:2:2}
For a constant main solenoid field, one can assume that the area $A$ and line density $\lambda_e$ of the electron beam stay constant. Following the same derivation as in Sec.~\ref{noiseamp:sec:1}, the electric field scales with the electron beam current and energy as (Eqn.~\ref{eqn:ecurrenscale}):
\begin{equation*}
	\mathbf{E}(E_{\mathrm{kin},e,2},I_{e,2})=\frac{I_{e,2}\beta_{e,1}}{I_{e,1}\beta_{e,2}}\mathbf{E}(E_{\mathrm{kin},e,1},I_1)
\end{equation*}
leading to
\begin{equation}
E_{\mathrm{center, integrated}}(10~\mathrm{keV},5~\mathrm{T},5.0~\mathrm{A})=105~\mathrm{kV}
\end{equation}
and an integrated kick of:
\begin{equation}
 \Delta x', \Delta y'= E_{\mathrm{center, integrated}}(10~\mathrm{keV},5~\mathrm{T},5.0~\mathrm{A})\cdot\frac{1}{B\rho_p\cdot v_p}=\q{15}{nrad}
\end{equation}

\section{Simulation of emittance growth and loss rates}
\label{sec:sim}
The simulations in preparation of the MD aim at identifying:
\begin{itemize}
	\item correlation between pulsing pattern and emittance growth and losses
	\item an estimate on the scaling of the emittance growth and losses with the kick amplitude
\end{itemize}
Based on earlier preliminary estimates, the integrated kick was estimated to be
\begin{equation}
\Delta x', \Delta y'= \q{12}{nrad}
\end{equation}
and the simulations presented in this paper are based on this value, explicitly \q{12}{nrad} and \q{120}{nrad} have been studied.

For the simulations, the lattice and optics are generated with MAD-X, SixTrack and SixDesk using the standard machinery available. The lattice is then imported into Lifetrac, with which the time evolution of the particle distribution is simulated over $10^6$ turns. As particle distribution, a 6D Gaussian distribution with $10^4$ macroparticles is used. In Sec.~\ref{sec:sim:1} we will first describe the lattice and optics generated with MAD-X and SixTrack and then present the Lifetrac simulation results in Sec.~\ref{sec:sim:2}.

\subsection{Lattice and optics preparation with SixTrack}
\label{sec:sim:1}
To represent the MD configuration, the 2016 injection optics, injection tunes and the standard chromaticity of $Q'_{x/y}=15$ and Landau damping octupole current of $I_{\rm MO}=\q{+19.6}{A}$ are used. The parameters are summarized in Table~\ref{sim:tab:1}.
\begin{table}[htb]
	\centering
	\caption{LHC 2016 optics parameters, injection optics. SKIPH and SKIPV are the positions where the resonant excitation is applied, which is close to the location of the ADT.\label{sim:tab:1}}
	\vspace{0.4cm}
	\begin{tabular}{llcc}
		\hline
		position & parameter & unit & value \\\hline\hline
		\multirow{3}{*}{IP1/5} & $\beta_{x/y}$ & m & 11.0/11.0 \\\hline
		& half crossing angle (alternated) & $\mu$rad & 170 \\\hline
		& half separation (alternated) & mm & 2 \\\hline
		SKIPH & position from IP3 & m & 3317.8 \\\hline
		& $\beta_{x/y}$ & m & 249.1/263.6 \\\hline
		SKIPV & position from IP3 & m & 3346.4 \\\hline
		 & $\beta_{x/y}$ & m & 222.8/285.2 \\\hline
	\end{tabular}
\end{table}

To study the influence of the magnetic errors, two different scenarios are considered:
\begin{itemize}
	\item \textbf{scenario 1 - no machine imperfections:} no magnetic errors
	\item \textbf{scenario 2 - including magnetic errors:} In RunII (2016) on average \q{1}{mm} rms orbit, 15\% peak $\beta$-beat are expected at injection \cite{rogelio}. The coefficients $a_1,b_1,a_2$ and $b_2$ of the multipolar field expansion of the magnetic errors are rescaled to deliver on average (over 60 seeds) \q{1}{mm} rms orbit, 15\% peak $\beta$-beat yielding on\_a1s=on\_a1r=on\_b1s=on\_b1r= 0.3 and on\_a2s=on\_a2r=on\_b2s=on\_b2r=1, where on\_* is a parameter for scaling the random ($r$) and systematic ($s$) errors assumed in the model. Otherwise standard errors $a_i$ and $b_i$ for $i\ge 3$ are assumed.
\end{itemize}
With this model, the average rms orbit and peak $\beta$-beat over all 60 seeds is then given by:
\begin{equation}
\rm{mean}\left(\rm{rms}\left(\Delta x/y\right)\right) = \q{1.05/1.36}{mm}, \quad \rm{mean}\left(\rm{max}\left(\frac{\Delta \beta_{x/y}}{\beta_{0,x/y}}\right)\right) = \q{15.94/13.49}{\%}
\end{equation}
which agrees well with requirement of \q{1}{mm} rms orbit and 15\% peak $\beta$-beat on average.
\begin{figure}[t]
	\centering
	\includegraphics[width=0.5\linewidth]{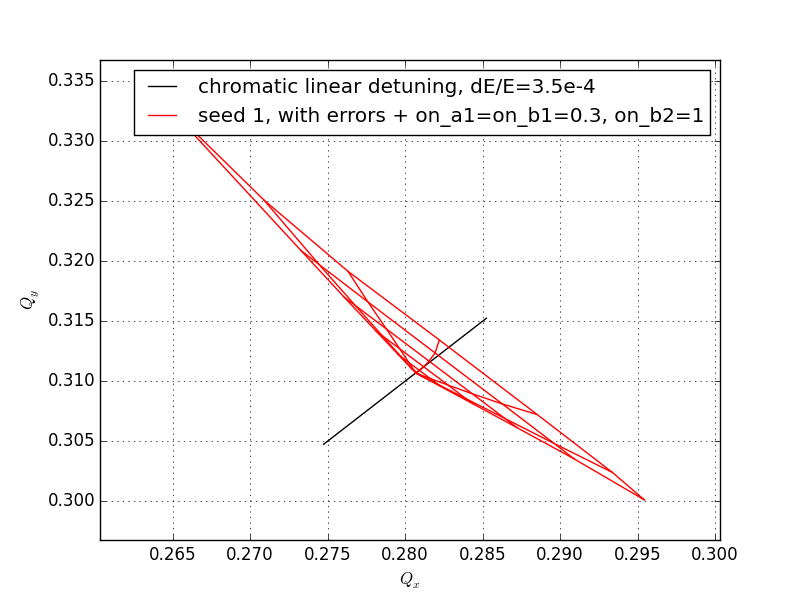}
	\caption{Tune footprint for RunII, 2016 injection optics, injection tunes, no collisions, $Q'_{x/y}=15$, $I_{\rm MO}=\q{+19.6}{A}$ and the seed used in the Lifetrac simulations (seed~1). Note that the tune footprint only varies very slightly between the different seeds.}
	\label{sim:fig:1}
\end{figure}

The resulting tune footprint for the seed used in the Lifetrac simulations (seed~1) is shown in Fig.~\ref{sim:fig:1}. Note that the tune footprint does not change considerably for the different seeds as the spread originates mainly from the Landau damping octupoles and the contribution from errors is small.
\begin{figure}[h!]
	\begin{minipage}[t]{1.0\linewidth}
		\centering
		\includegraphics[width=0.8\linewidth]{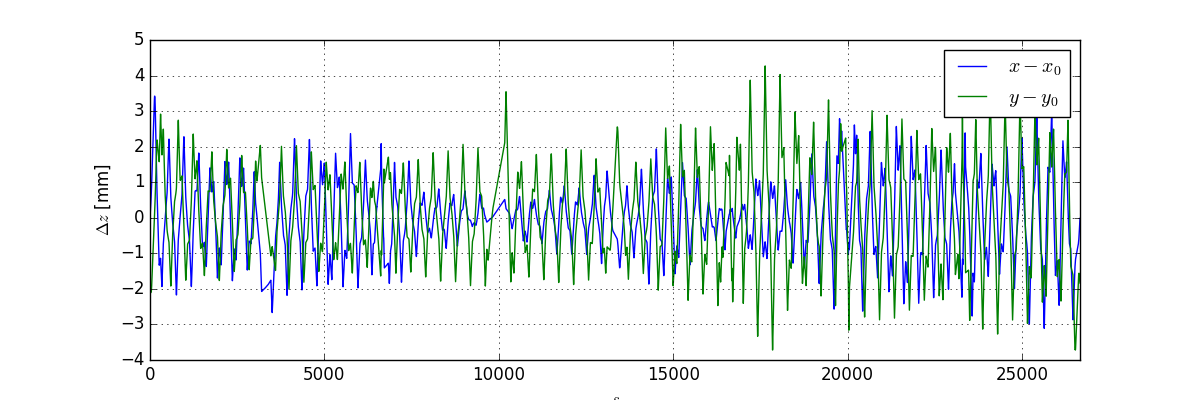}
	\end{minipage}
	\begin{minipage}[t]{1.0\linewidth}
		\centering
		\includegraphics[width=0.8\linewidth]{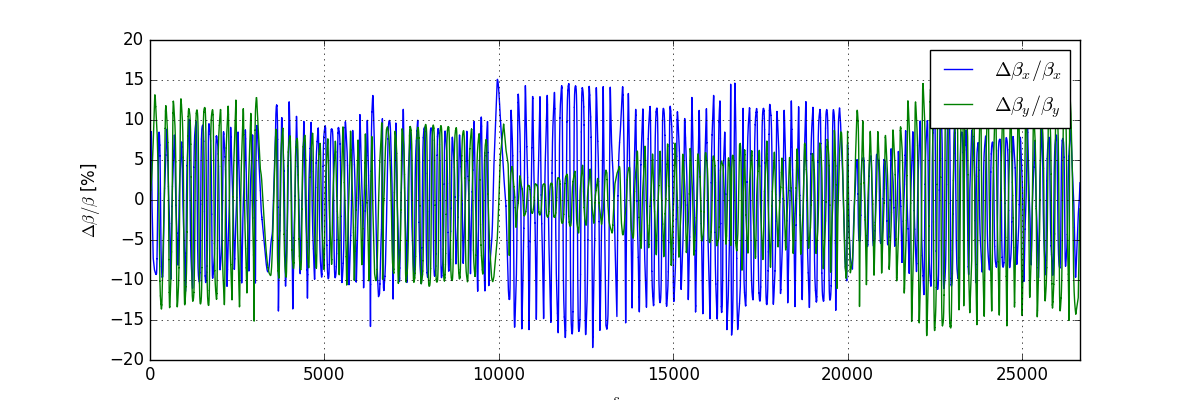}
	\end{minipage}
	\caption{Hor. and Vert. closed orbit (top) and $\beta$-beat (bottom) for the seed used in the Lifetrac simulations (seed~1). 2016 injection optics, injection tunes, no collisions, $Q'_{x/y}=15$ and $I_{\rm MO}=\q{+19.6}{A}$ are assumed.  The plots show the deviation of in the horizontal (blue) and vertical (green) plane from the case without errors. The sequence starts at IP3.}
	\label{sim:fig:2}
\end{figure}
The rms orbit and peak $\beta$-beat of the seed used in the Lifetrac simulations (seed~1) is:
\begin{equation}
\rm{rms}\left(\Delta x/y\right) = \q{0.73/1.34}{mm}, \quad \rm{max}\left(\frac{\Delta \beta_{x/y}}{\beta_{0,x/y}}\right) = \q{12.03/10.11}{\%}
\end{equation}
and the variation along the machine of the orbit and $\beta$-beat for this seed is illustrated in Fig.~\ref{sim:fig:2}.

\subsection{Lifetrac simulation results}
\label{sec:sim:2}
To study the time evolution of the emittance, a 6D~Gaussian distribution with $10^4$ macroparticles is tracked over $10^6$ turns in steps of $10^4$~turns. To reduce the statistical error, emittance and bunch length are averaged over $10^4$ turns under the assumption that changes are minimal within this time span. The ADT excitation is simulated as a horizontal kicker inserted before the RF cavity ACSCA.D5L4.B1 and a vertical kicker after the RF cavity ACSCA.D5R4.B1 at approximately 12~m to 14~m from IP4. This position corresponds roughly to the position of the horizontal/vertical kicker of the ADT in the LHC. Furthermore, only Beam~1 is simulated, similar results are expected for Beam~2.

This section is divided into two parts. First the simulation results for the optics without any machine imperfections is presented (Sec.~\ref{sec:sim:2:1}), then the ones including the standard magnetic errors and 15\% average peak $\beta$-beat and 1~mm average rms orbit (Sec.~\ref{sec:sim:2:2}) in order to study the influence of the magnetic errors.

\subsubsection{Scenario 1: No machine imperfections}
\label{sec:sim:2:1}
\begin{figure}[h!]
	\begin{minipage}[t]{0.49\linewidth}
		\centering
		\includegraphics[width=1.0\linewidth]{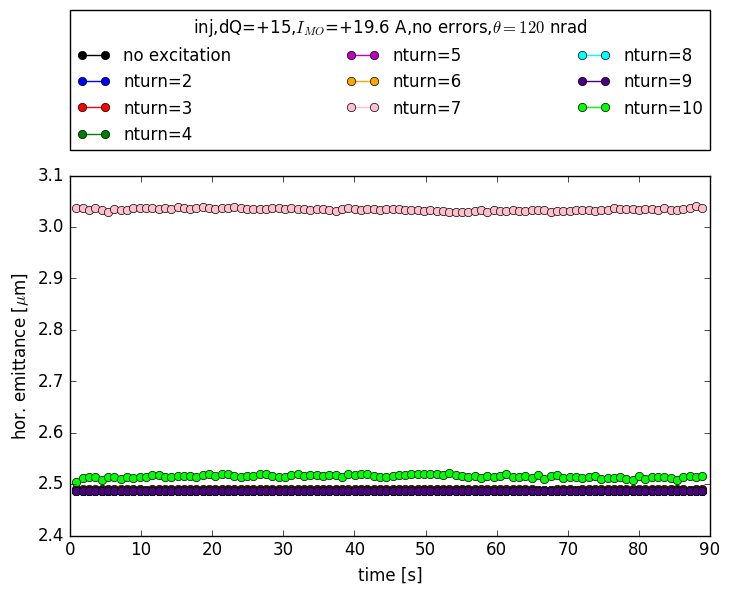}
	\end{minipage}
	\begin{minipage}[t]{0.49\linewidth}
		\centering
		\includegraphics[width=1.0\linewidth]{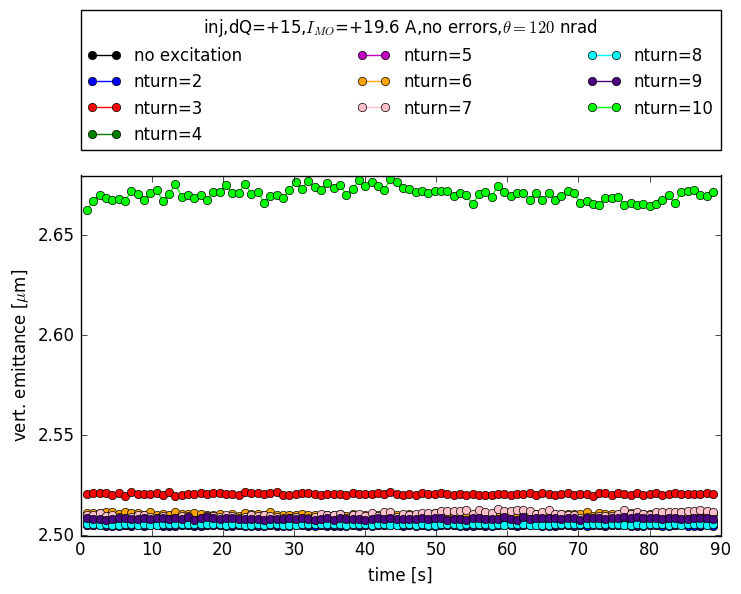}
	\end{minipage}
	\begin{minipage}[t]{0.49\linewidth}
		\centering
		\includegraphics[width=1.0\linewidth]{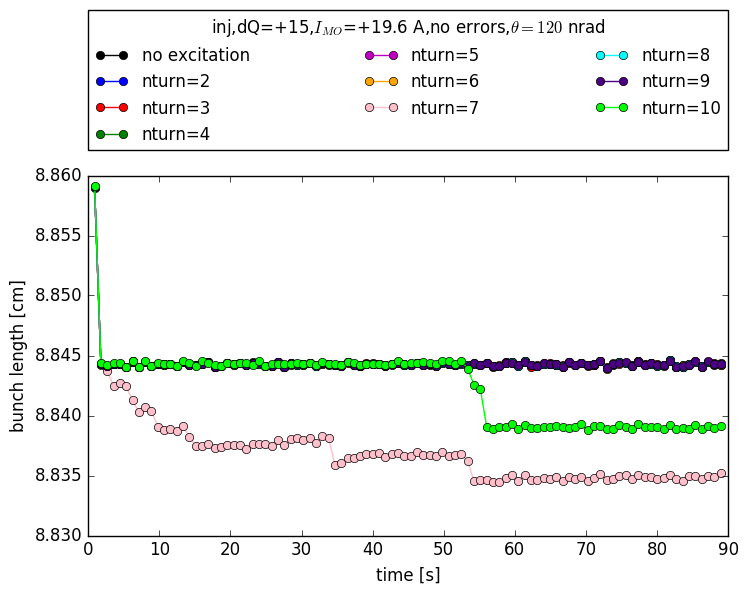}
	\end{minipage}
	\begin{minipage}[t]{0.49\linewidth}
		\centering
		\includegraphics[width=1.0\linewidth]{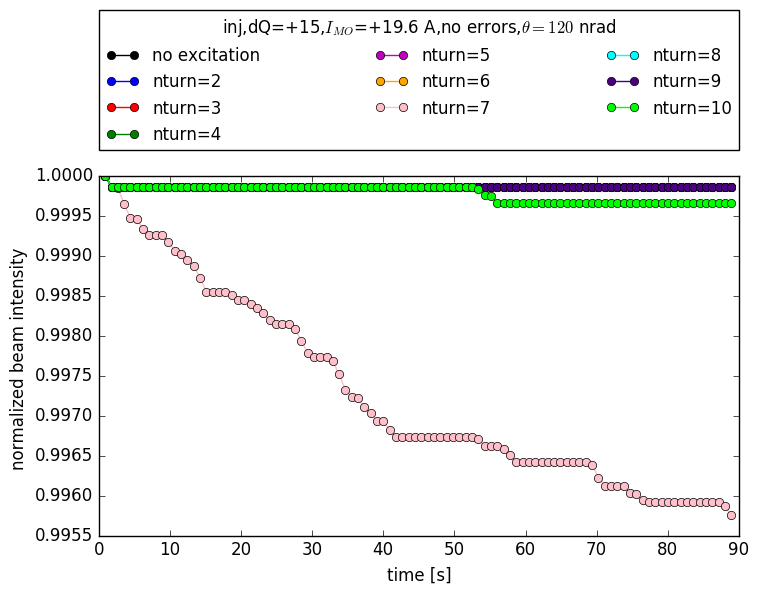}
	\end{minipage}
	\caption{Scenario 1 (no errors): Hor. (top left) and Vert. (top right) normalized emittance and $1\sigma$ rms bunch length (bottom left) and normalized beam intensity (bottom right) over $10^6$ turns. A change in emittance is observed for pulsing every 7th and 10th turn and to a much smaller extent every 3rd turn. The change in beam distribution for pulsing every 7th and 10th turn takes place over the first $10^4$ turns. Changes of both are only observed for pulsing every 7th and every 10th turn. Intensity loss accompanied by bunch shortening indicates longitudinal losses. The step in the bunch length during the first turn is due to the slight mismatch of the distribution to the non-linear RF bucket. \label{sim:fig:2:1:1}}
\end{figure}

For the simulations the 2016 injection optics with chromaticity and Landau damping octupole settings of $Q'_{x/y}=15$, $I_{\rm MO}=\q{+19.6}{A}$ are used as described in detail in Sec.~\ref{sec:sim}, Scenario 1. The transverse emittance, bunch length and normalized beam intensity for an excitation amplitude of $120$~nrad and different pulsing patterns are shown in Fig.~\ref{sim:fig:2:1:1}. A change in emittance, bunch length or occurrence of losses is observed for pulsing every 7th and 10th turn and to a much smaller extent every 3rd turn.

\begin{figure}[t]
	\begin{minipage}[t]{0.49\linewidth}
		\centering
		\includegraphics[width=1.0\linewidth]{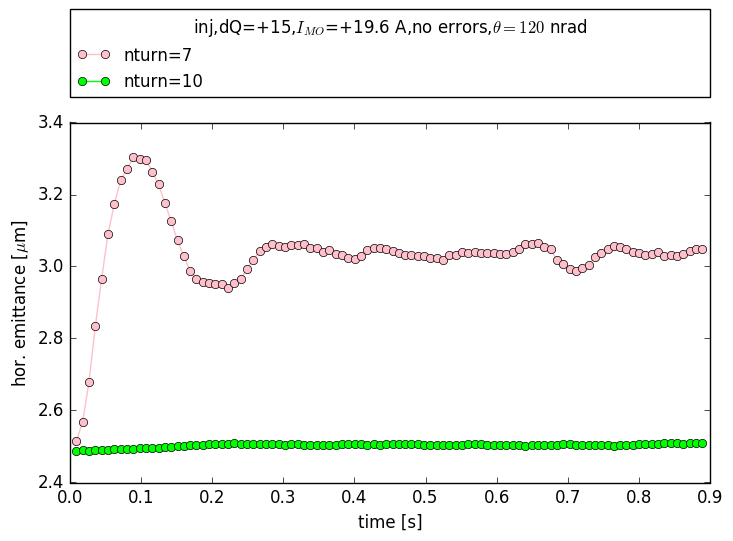}
	\end{minipage}
	\begin{minipage}[t]{0.49\linewidth}
		\centering
		\includegraphics[width=1.0\linewidth]{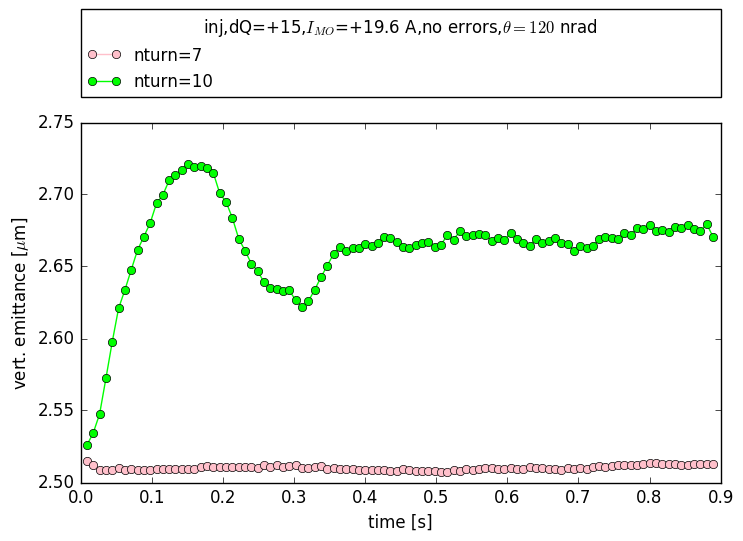}
	\end{minipage}
	\caption{Scenario 1 (no errors), excitation amplitude 120~nrad: Hor. (left) and Vert. (right) normalized emittance over $10^4$~turns. The change in emittance indicates the change in distribution happening over the first $10^4$~turns.\label{sim:fig:2:1:3}}
\end{figure}
\begin{figure}[h]
	\begin{minipage}[t]{0.32\linewidth}
		\centering
		\hbox{}
		\includegraphics[width=1.0\linewidth]{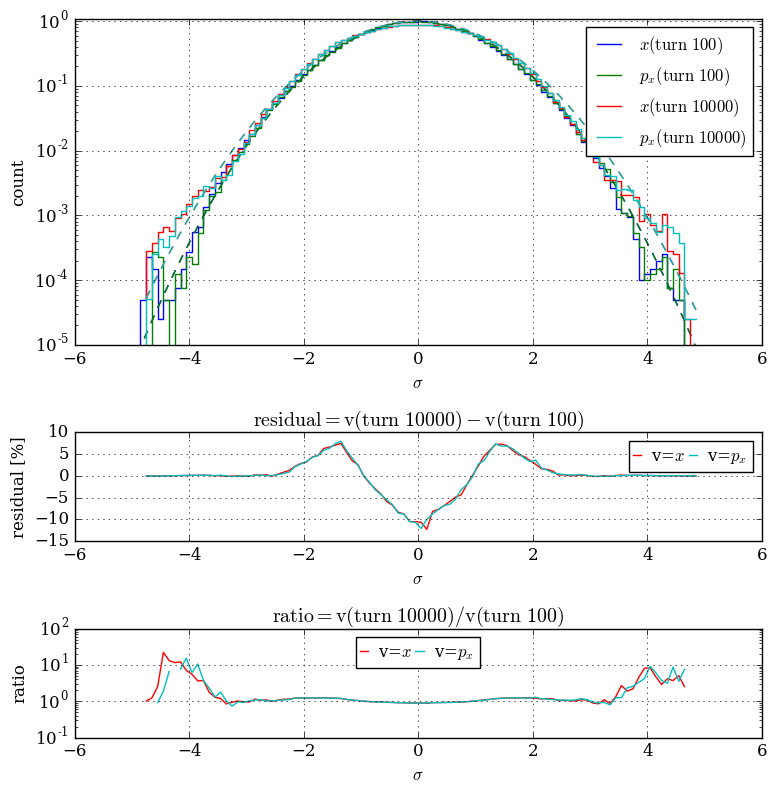}
	\end{minipage}
	\begin{minipage}[t]{0.32\linewidth}
		\centering
		7th turn pulsing
		\includegraphics[width=1.0\linewidth]{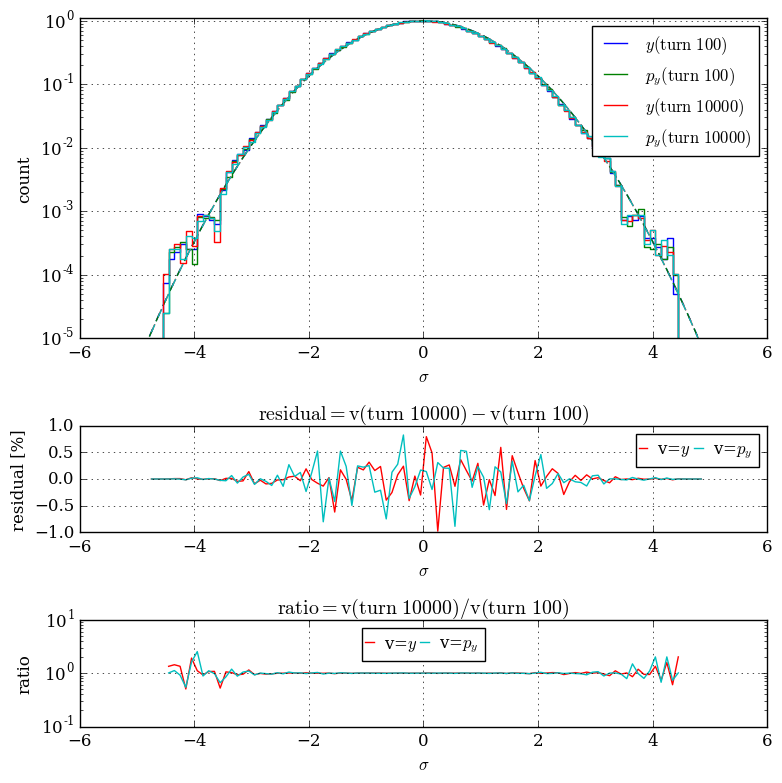}
	\end{minipage}
	\begin{minipage}[t]{0.32\linewidth}
		\centering
		\hbox{}
		\includegraphics[width=1.0\linewidth]{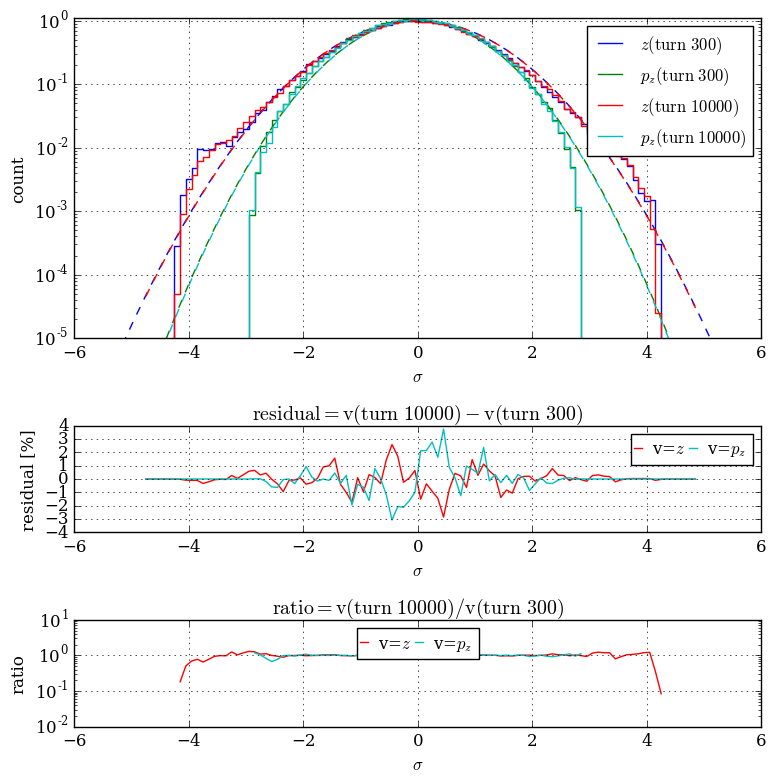}
	\end{minipage}
	\begin{minipage}[t]{0.32\linewidth}
		\centering
		\hbox{}
		\includegraphics[width=1.0\linewidth]{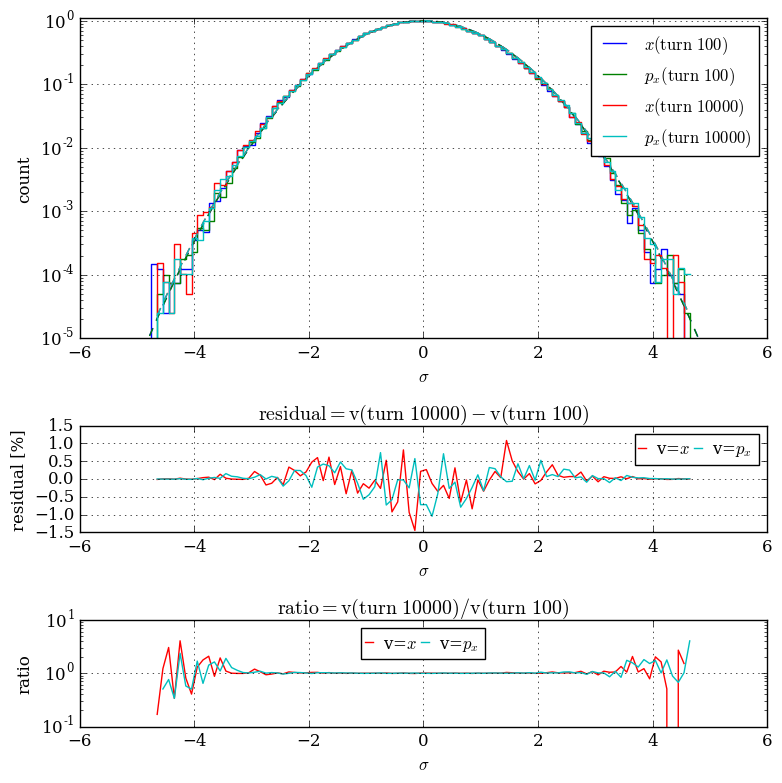}
	\end{minipage}
	\begin{minipage}[t]{0.32\linewidth}
		\centering
		10th turn pulsing
		\includegraphics[width=1.0\linewidth]{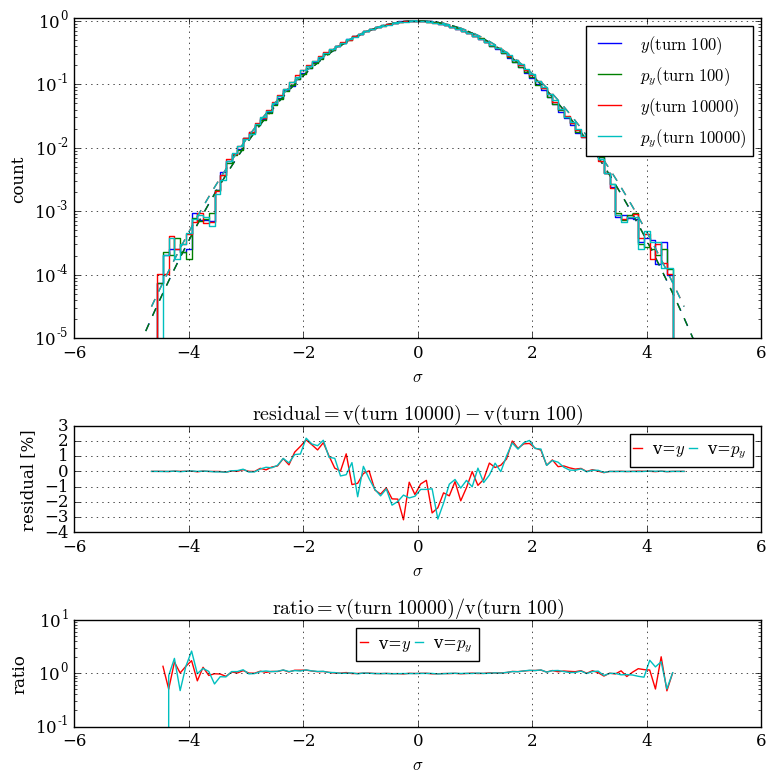}
	\end{minipage}
	\begin{minipage}[t]{0.32\linewidth}
		\centering
		\hbox{}
		\includegraphics[width=1.0\linewidth]{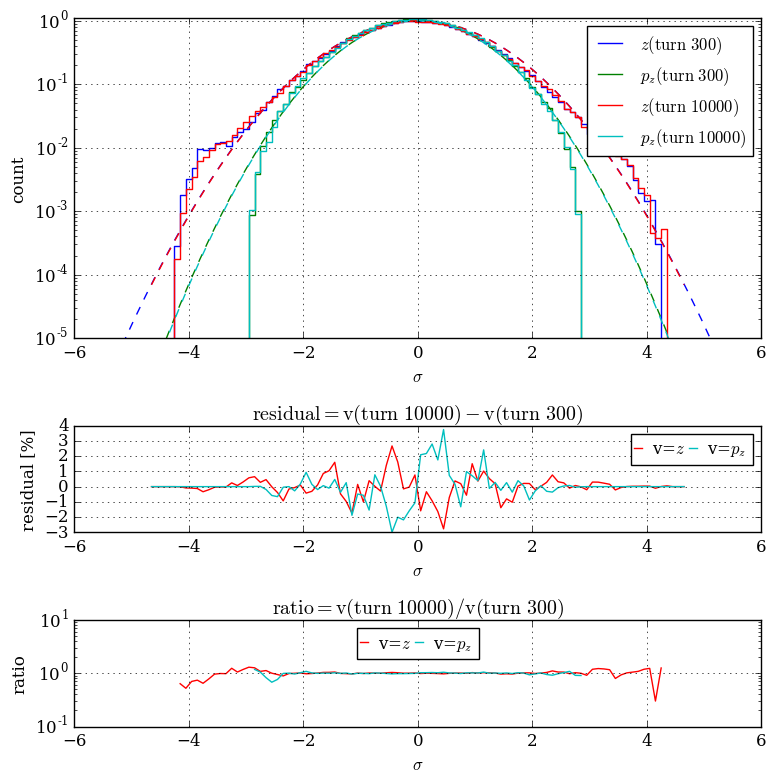}
	\end{minipage}
	\caption{Scenario 1 (no errors), excitation amplitude 120~nrad: Normalized amplitude distribution in $(x,p_x)$ (left), $(y,p_y)$ (center) and $(z,p_z)$ (right) for pulsing every 7th (top) and 10th turn (bottom) with an amplitude of 120~nrad. The change in distribution is too small to be directly detected in the histograms.\label{sim:fig:2:1:5}}
\end{figure}

Both cases exhibit a fast increase in hor. (7th turn) or vert. (10th turn) emittance and later longitudinal losses expressed by a decrease in intensity \emph{and} bunch length. The fast increase in emittance is in fact an adjustment of the beam distribution over $10^4$ turns as confirmed by the changing emittance over the first $10^4$~turns (Fig.~\ref{sim:fig:2:1:3}) and by comparison with the initial distribution and distribution after $10^4$~turns shown in Fig.~\ref{sim:fig:2:1:5}. For pulsing every 7th turn the horizontal distribution decreases in the center of the distribution and increase around $1.5\ \sigma$. Furthermore also a visible increase of the tails is observed around $4\ \sigma$. For pulsing every 10th turn, the distribution in the center decreases as well and increases around $2\ \sigma$. A statistically significant increase for higher amplitudes is not observed.

The FMA analysis further gives information about the excited resonances (Appendix~\ref{appendix:1}, Fig.~\ref{sim:fig:2:1:6}). For pulsing every 7th turn, the $7Q_x=2$ resonance is driven leading to a blow-up in the horizontal plane. As octupoles only drive even resonances, this resonance originates from the strong sextupoles, while the octupoles role was to generate the large tune footprint. The large chromaticity then leads to a repeated crossing of the core particles over the $7Q_x$ resonances due to the synchrotron motion and chromatic detuning, which explains the blow-up of the core and the mainly longitudinal losses. Without a high chromaticity only losses in the transverse tails of the distribution would be expected (see FMA anlysis in amplitude space Appendix~\ref{appendix:1}, Fig.~\ref{sim:fig:2:1:7}). For pulsing every 10th turn, the $10Q_y$ or $10Q_x$ resonances are excited. The change in emittance in the vertical but not horizontal plane indicates that only the $10Q_y$ resonance is excited. Losses occur again longitudinally as the particles are repeatedly approaching or crossing the resonance because of the synchrotron motion. 
\clearpage

\subsubsection{Scenario 2: Including standard magnetic errors, 15\% average peak $\beta$-beat and 1~mm average rms orbit}
\label{sec:sim:2:2}
\begin{figure}[h]
	\begin{minipage}[t]{0.49\linewidth}
		\centering
		\includegraphics[width=1.0\linewidth]{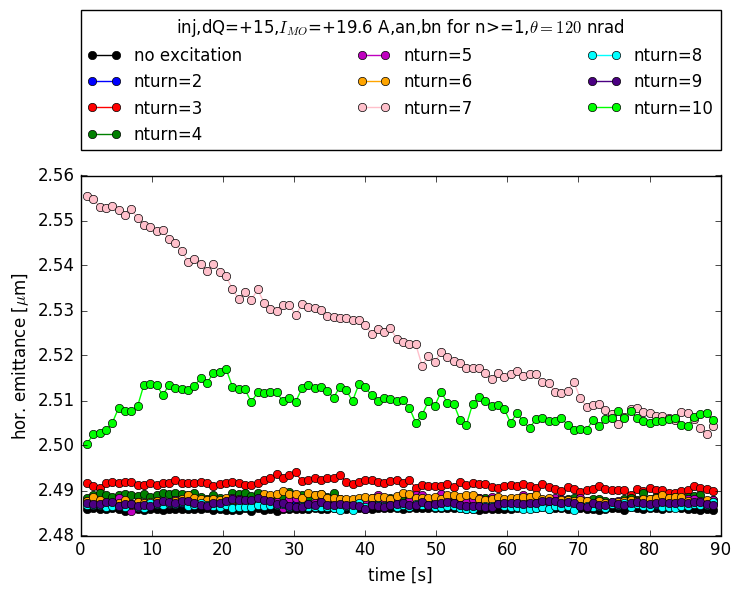}
	\end{minipage}
	\begin{minipage}[t]{0.49\linewidth}
		\centering
		\includegraphics[width=1.0\linewidth]{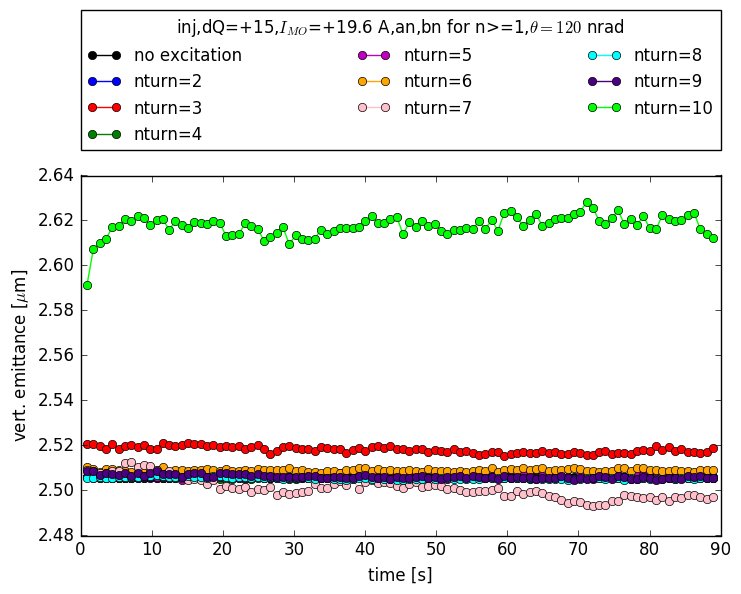}
	\end{minipage}
	\begin{minipage}[t]{0.49\linewidth}
		\centering
		\includegraphics[width=1.0\linewidth]{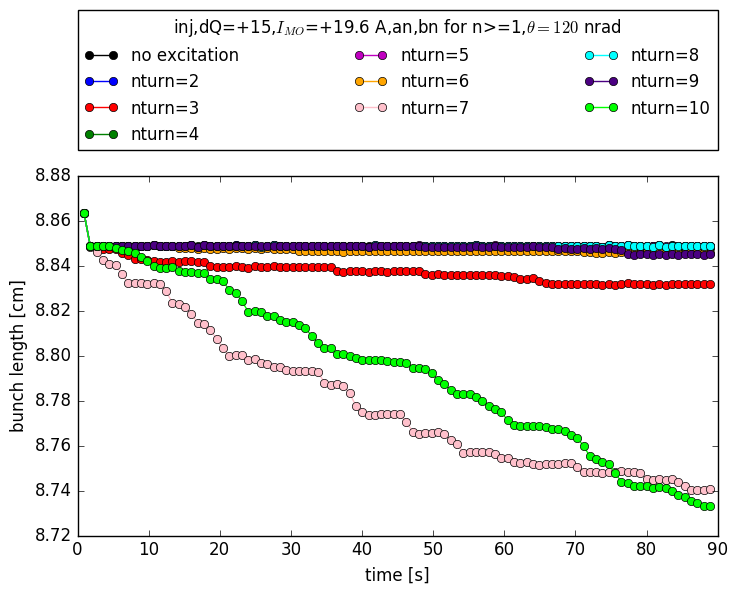}
	\end{minipage}
	\begin{minipage}[t]{0.49\linewidth}
		\centering
		\includegraphics[width=1.0\linewidth]{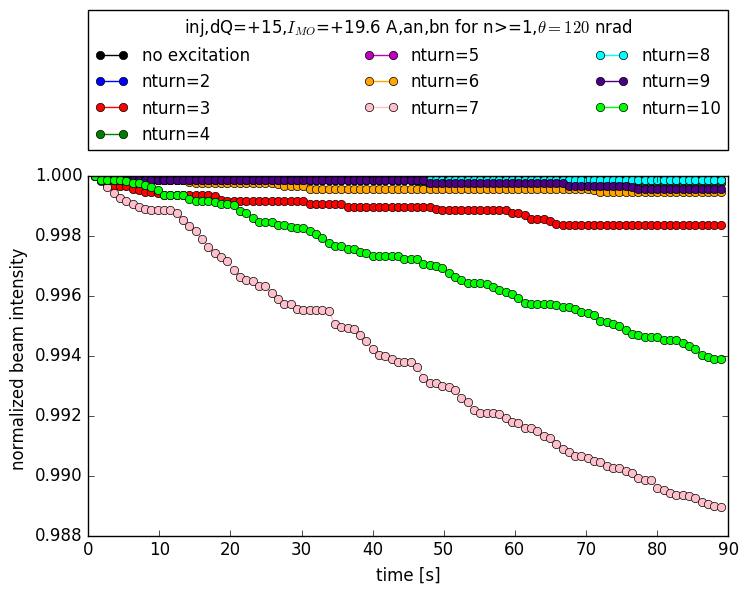}
	\end{minipage}
	\caption{Scenario 2 (with machine imperfections), excitation amplitude 120~nrad: Hor. (top left) and Vert. (top right) normalized emittance and $1\sigma$ rms bunch length (bottom left) and normalized beam intensity (bottom right) over $10^6$ turns. A change in emittance is only observed for pulsing every 7th and 10th turn. This change in beam distribution takes place over the first $10^4$ turns. Changes of both are mainly observed for pulsing every 7th and every 10th turn and to a much smaller for the other pulsing patterns. As the bunch length decreases as also the beam intensity decreases, the beam losses are longitudinal. \label{sim:fig:2:2:1}}
\end{figure}

To study the impact of machine imperfections, Scenario~2 is a copy of Scenario~1 including realistic machine imperfections, explicitly the standard magnetic errors for $a_i,b_i, i\ge 2$ and 30\% of the standard errors for $a_1$and $b_1$ in order to obtain around 15\% average peak $\beta$-beat and 1~rms average orbit (Sec.~\ref{sec:sim}, Scenario 2). The transverse emittance, bunch length and normalized beam intensity for an excitation amplitude of $120$~nrad and different pulsing patterns are shown in Fig.~\ref{sim:fig:2:2:1}.

In comparison to Scenario~1 without errors, the following observations can be made:
\begin{itemize}
\item Losses occur now also for pulsing patterns other than pulsing every 7th and 10th turn due to the magnetic errors containing up to 14th order in $a_1$ and $b_1$. However, the strongest losses are still observed for pulsing every 7th and 10th turn indicating that the resonances are generated and driven by an interplay of sextupoles and octupoles as already present in the case without magnetic imperfections (Scenario~1)
\item Considerable changes in emittance are only observed for pulsing every 3rd, 7th and 10th turn, where the emittance growth for pulsing every 3rd turn is comparatively small.
\item The decrease in emittance for pulsing every 7th and 10th turn indicates that the losses are not only longitudinal as in the case without errors, but also transverse.
\end{itemize}
The results of the FMA analysis for the different pulsing patterns are shown in Appendix~\ref{appendix:2}.

In preparation of the beam studies, it is furthermore important to know the dependence of the time evolution of the emittance, bunch length and beam losses on the excitation amplitude. A decrease of the excitation amplitude by a factor~10, namely an excitation amplitude of $12$~nrad, leads to stable beams over $10^6$ turns for all pulsing patterns (Appendix~\ref{appendix:3}, Fig.~\ref{sim:fig:2:2:3}--\ref{sim:fig:2:2:4})

\section{Conclusions}

In preparation of beam studies in the LHC on the effect of a resonant excitation on the beam core, simulations of the experimental scenario (2016 injection optics, injection tunes, separated beams, $Q_{x/y}'=+15$ and Landau damping octupoles at $I_{\rm MO}=19.6$~A.) and different pulsing patterns and amplitudes have been performed.

Without errors, an effect of the pulsing is only observed for pulsing every 7th and 10th turn in terms of:
\begin{itemize}
	\item longitudinal losses
	\item an adjustment of the beam distribution over $10^4$ turns to a steady distribution with increased emittance
\end{itemize}  
That an effect is only present for pulsing every 7th and 10th turn can be explained by an excitation of the 7th and 10th order resonances driven by the strong sextupoles. The (mainly longitudinal) losses are due to the high chromaticity, as the synchrotron motion and the chromatic detuning lead to a repeated crossing of the off-momentum particles over the resonances.

With machine imperfections, the strongest effect is observed for pulsing every 7th and 10th turn indicating that the main effect originates from the strong sextupoles and octupoles and high chromaticity. The losses are now not only longitudinal, but also transverse observable as a decrease in transverse emittance \emph{and} bunch length. In addition, also other excitation patterns ---mainly every 3rd turn--- show losses due to the magnetic errors present in simulations up to 14th order.

Most of the calculations were done with an excitation amplitude of 120~nrad. A reduction of the amplitude by a factor~10, i.e. 12~nrad, shows no observable effects on the beam core, even with machine imperfections.

\begin{acknowledgments}

We would like to thank D.~Shatilov for his support with Lifetrac, S. Fartoukh, R. De Maria, R. Tom\'{a}s and Y.~Papaphilippou for their invaluable help with the preparation of the lattice and optics input file for SixTrack (mask file) and R.~Bruce and S. Redaelli for their advice with the simulation parameters.

\end{acknowledgments}
\clearpage
\appendix
\section{Scenario 1 (no errors): FMA analysis for 120~nrad excitation amplitude}
\label{appendix:1}
\begin{figure}[h]
	\begin{minipage}[t]{0.3\linewidth}
		\centering
		no excitation
		\includegraphics[width=1.0\linewidth]{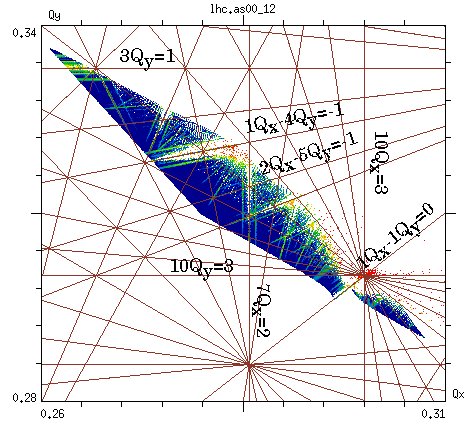}
	\end{minipage}
	\begin{minipage}[t]{0.3\linewidth}
		\centering
		7th turn pulsing
		\includegraphics[width=1.0\linewidth]{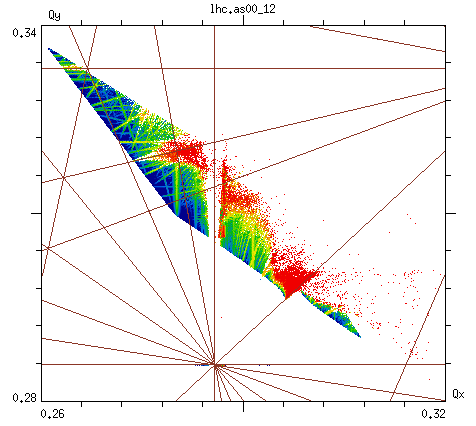}
	\end{minipage}
	\begin{minipage}[t]{0.3\linewidth}
		\centering
		10th turn pulsing
		\includegraphics[width=1.0\linewidth]{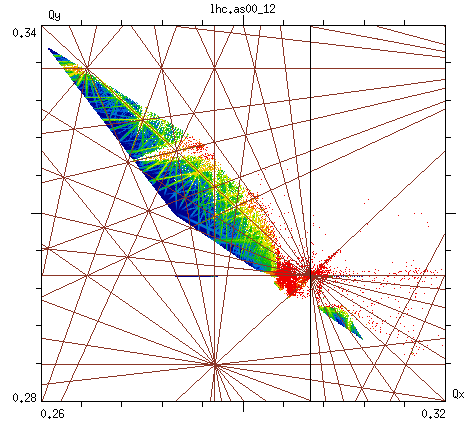}
	\end{minipage}
	\caption{Scenario 1 (no errors): FMA analysis for on-momentum particles ($\frac{\Delta p}{p_0}=0$) up to $8~\sigma$ amplitude for a square grid and 120~nrad excitation amplitude: no excitation (left), pulsing every 7th turn (center) and pulsing every 10th turn (right).\label{sim:fig:2:1:6}}
\end{figure}

\begin{figure}[h]
	\begin{minipage}[t]{0.33\linewidth}
		\centering
		no excitation
		\includegraphics[width=1.0\linewidth]{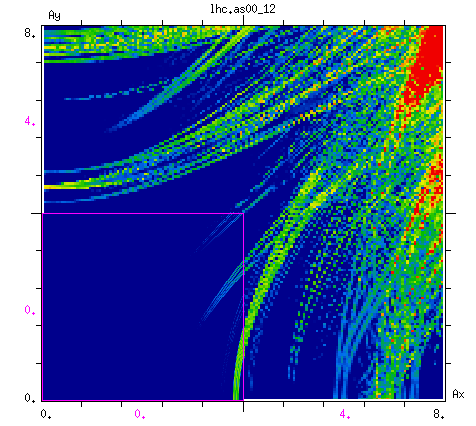}
	\end{minipage}
	\begin{minipage}[t]{0.33\linewidth}
		\centering
		7th turn pulsing
		\includegraphics[width=1.0\linewidth]{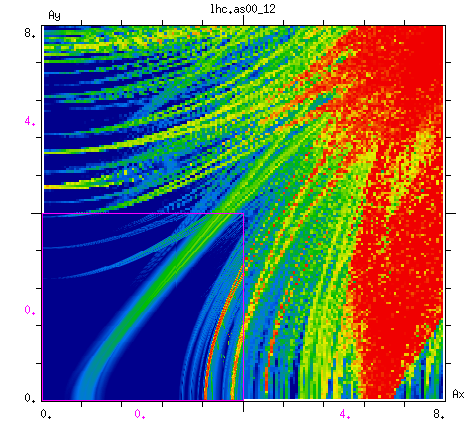}
	\end{minipage}
	\begin{minipage}[t]{0.33\linewidth}
		\centering
		10th turn pulsing
		\includegraphics[width=1.0\linewidth]{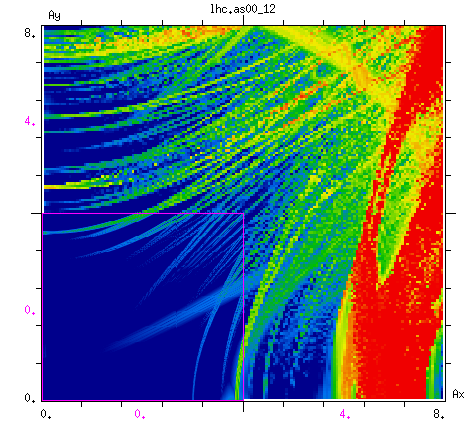}
	\end{minipage}
	\begin{minipage}[t]{0.33\linewidth}
		\centering
		no excitation
		\includegraphics[width=1.0\linewidth]{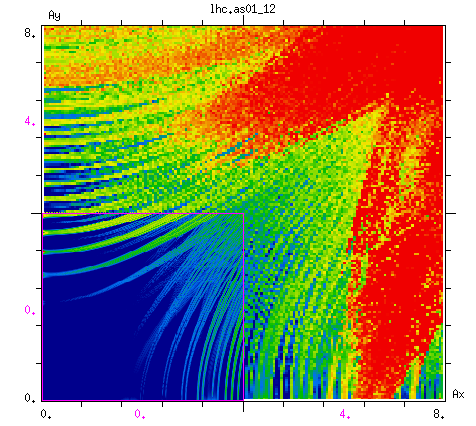}
	\end{minipage}
	\begin{minipage}[t]{0.33\linewidth}
		\centering
		7th turn pulsing
		\includegraphics[width=1.0\linewidth]{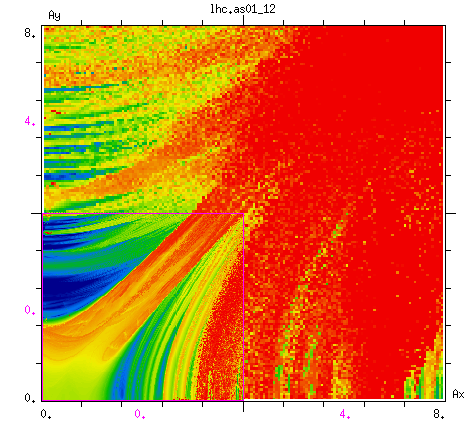}
	\end{minipage}
	\begin{minipage}[t]{0.33\linewidth}
		\centering
		10th turn pulsing
		\includegraphics[width=1.0\linewidth]{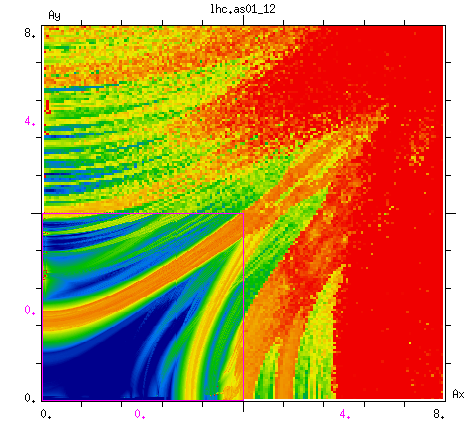}
	\end{minipage}
	\caption{Scenario 1 (no errors): FMA analysis in amplitude space for on-momentum particles ($\frac{\Delta p}{p_0}=0$) (top) and off-momentum ($\frac{\Delta p}{p_0}=0.1 \sigma_p$) (bottom) up to $8~\sigma$ amplitude for a square grid and 120~nrad excitation amplitude: no excitation (left), pulsing every 7th turn (center) and pulsing every 10th turn (right).\label{sim:fig:2:1:7}}
\end{figure}

\clearpage
\section{Scenario 2 (with magnetic imperfections): FMA analysis for different pulsing patterns and excitation amplitude of 120~nrad}
\label{appendix:2}
\begin{figure}[h]
	\begin{minipage}[t]{0.28\linewidth}
		\centering
		no excitation
		\includegraphics[width=1.0\linewidth]{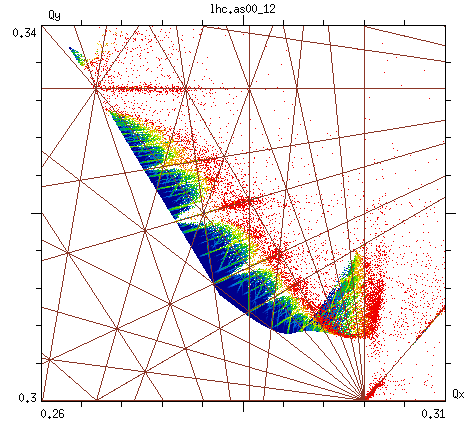}
	\end{minipage}
	\begin{minipage}[t]{0.28\linewidth}
		\centering
		7th turn pulsing
		\includegraphics[width=1.0\linewidth]{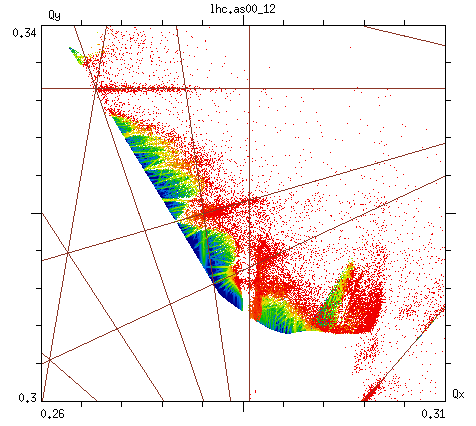}
	\end{minipage}
	\begin{minipage}[t]{0.28\linewidth}
		\centering
		10th turn pulsing
		\includegraphics[width=1.0\linewidth]{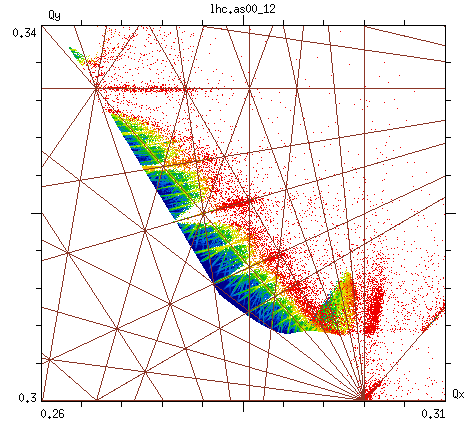}
	\end{minipage}
	\begin{minipage}[t]{0.28\linewidth}
		\centering
		2nd turn pulsing
		\includegraphics[width=1.0\linewidth]{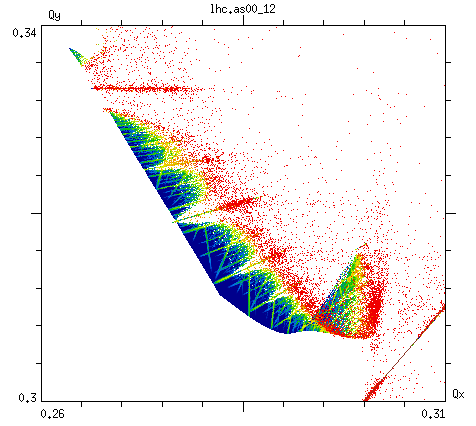}
	\end{minipage}
	\begin{minipage}[t]{0.28\linewidth}
		\centering
		3rd turn pulsing
		\includegraphics[width=1.0\linewidth]{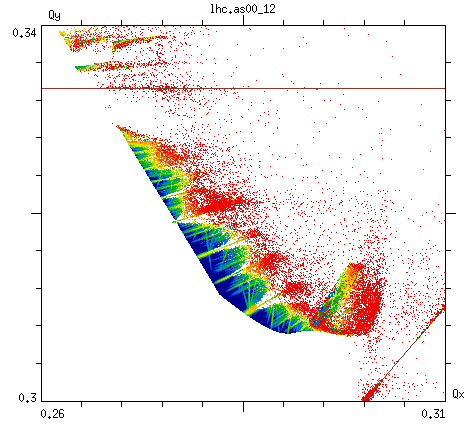}
	\end{minipage}
	\begin{minipage}[t]{0.28\linewidth}
		\centering
		4th turn pulsing
		\includegraphics[width=1.0\linewidth]{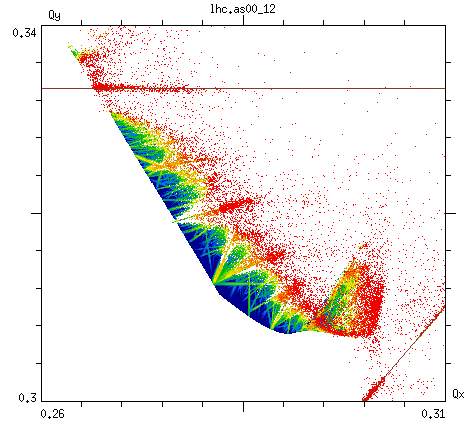}
	\end{minipage}
	\begin{minipage}[t]{0.28\linewidth}
		\centering
		5th turn pulsing
		\includegraphics[width=1.0\linewidth]{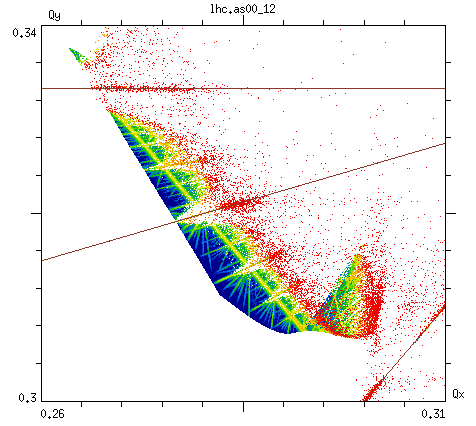}
	\end{minipage}
	\begin{minipage}[t]{0.28\linewidth}
		\centering
		6th turn pulsing
		\includegraphics[width=1.0\linewidth]{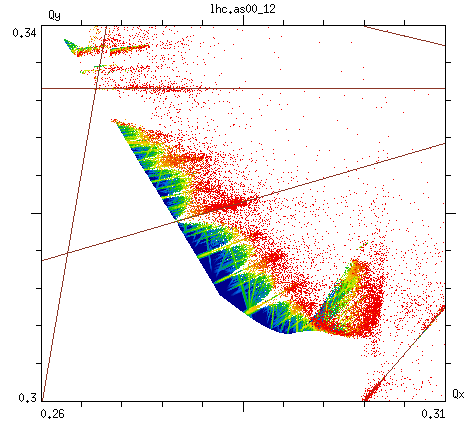}
	\end{minipage}
	\begin{minipage}[t]{0.28\linewidth}
		\centering
		8th turn pulsing
		\includegraphics[width=1.0\linewidth]{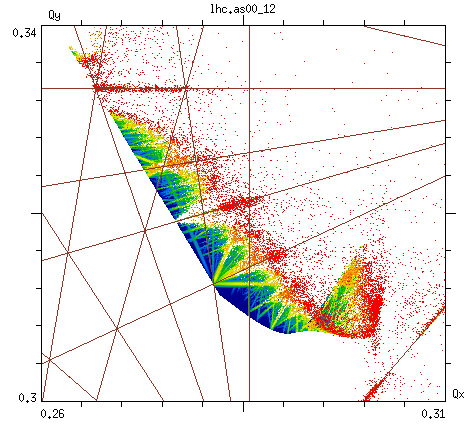}
	\end{minipage}
	\begin{minipage}[t]{0.28\linewidth}
		\centering
		9th turn pulsing
		\includegraphics[width=1.0\linewidth]{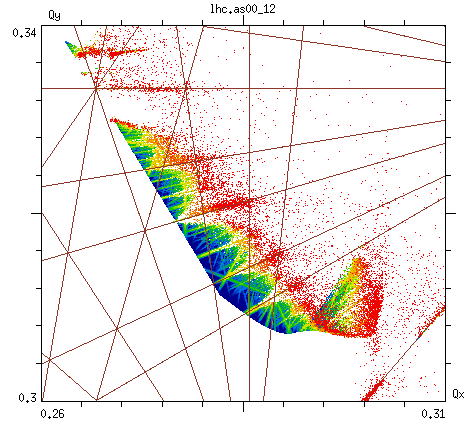}
	\end{minipage}
	\caption{Scenario 2 (with magnetic imperfections): FMA analysis for on-momentum particles ($\frac{\Delta p}{p_0}=0.0 \sigma_p$) up to $8~\sigma$ amplitude for a square grid and 120~nrad excitation amplitude and different pulsing patterns. The resonance lines up to the same order as the pulsing pattern (e.g. 10th turn pulsing up to 10th order resonance) are indicated in red.\label{sim:fig:2:2:7}}
\end{figure}

\begin{figure}[h]
	\begin{minipage}[t]{0.3\linewidth}
		\centering
		no excitation
		\includegraphics[width=1.0\linewidth]{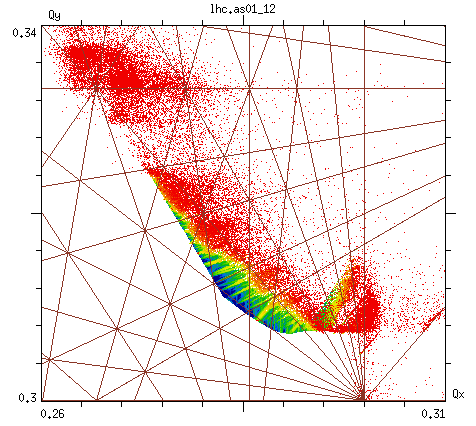}
	\end{minipage}
	\begin{minipage}[t]{0.3\linewidth}
		\centering
		7th turn pulsing
		\includegraphics[width=1.0\linewidth]{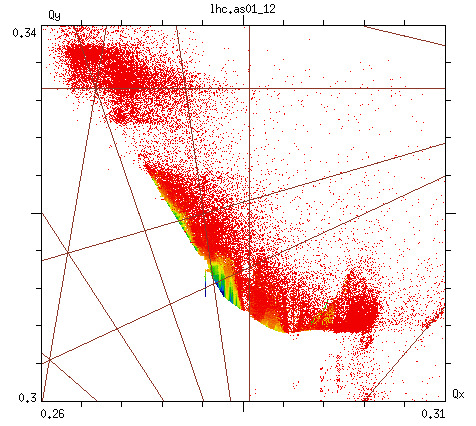}
	\end{minipage}
	\begin{minipage}[t]{0.3\linewidth}
		\centering
		10th turn pulsing
		\includegraphics[width=1.0\linewidth]{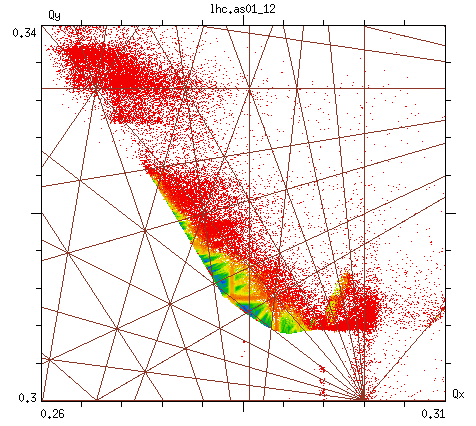}
	\end{minipage}
	\begin{minipage}[t]{0.3\linewidth}
		\centering
		2nd turn pulsing
		\includegraphics[width=1.0\linewidth]{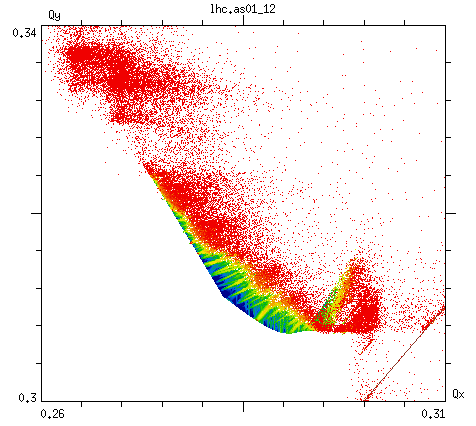}
	\end{minipage}
	\begin{minipage}[t]{0.3\linewidth}
		\centering
		3rd turn pulsing
		\includegraphics[width=1.0\linewidth]{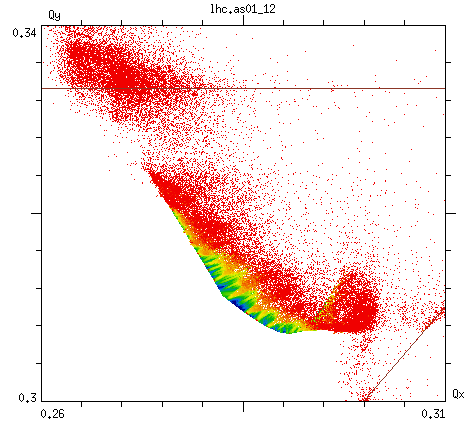}
	\end{minipage}
	\begin{minipage}[t]{0.3\linewidth}
		\centering
		4th turn pulsing
		\includegraphics[width=1.0\linewidth]{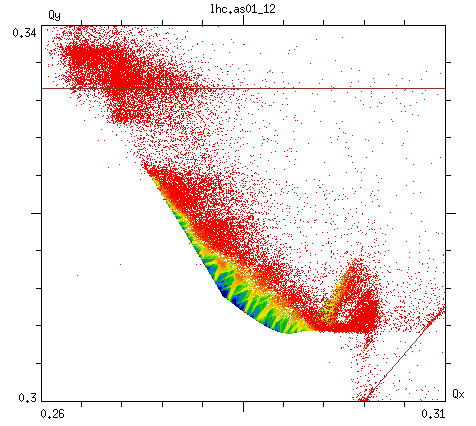}
	\end{minipage}
	\begin{minipage}[t]{0.3\linewidth}
		\centering
		5th turn pulsing
		\includegraphics[width=1.0\linewidth]{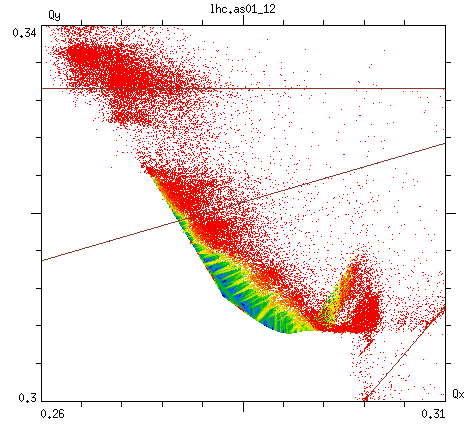}
	\end{minipage}
	\begin{minipage}[t]{0.3\linewidth}
		\centering
		6th turn pulsing
		\includegraphics[width=1.0\linewidth]{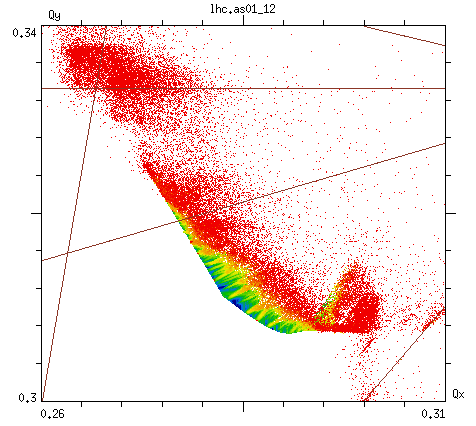}
	\end{minipage}
	\begin{minipage}[t]{0.3\linewidth}
		\centering
		8th turn pulsing
		\includegraphics[width=1.0\linewidth]{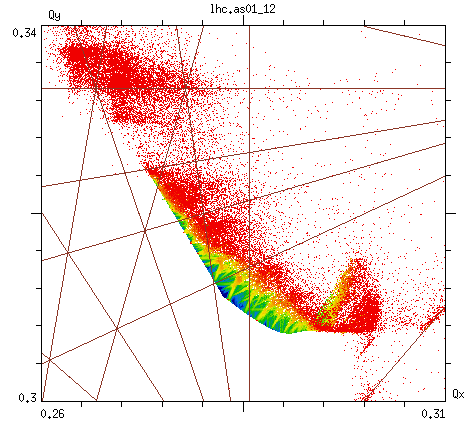}
	\end{minipage}
	\begin{minipage}[t]{0.3\linewidth}
		\centering
		9th turn pulsing
		\includegraphics[width=1.0\linewidth]{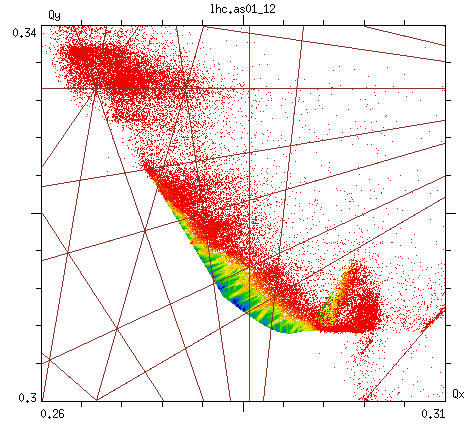}
	\end{minipage}
	\caption{Scenario 2 (with magnetic imperfections): FMA analysis for off-momentum particles ($\frac{\Delta p}{p_0}=0.1 \sigma_p$) up to $8~\sigma$ amplitude for a square grid and 120~nrad excitation amplitude and different pulsing patterns. The resonance lines up to the same order as the pulsing pattern (e.g. 10th turn pulsing up to 10th order resonance) are indicated in red.\label{sim:fig:2:2:8}}
\end{figure}
\begin{figure}[h]
	\begin{minipage}[t]{0.3\linewidth}
		\centering
		no excitation
		\includegraphics[width=1.0\linewidth]{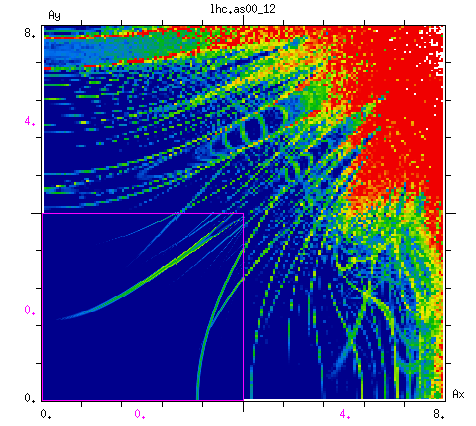}
	\end{minipage}
	\begin{minipage}[t]{0.3\linewidth}
		\centering
		7th turn pulsing
		\includegraphics[width=1.0\linewidth]{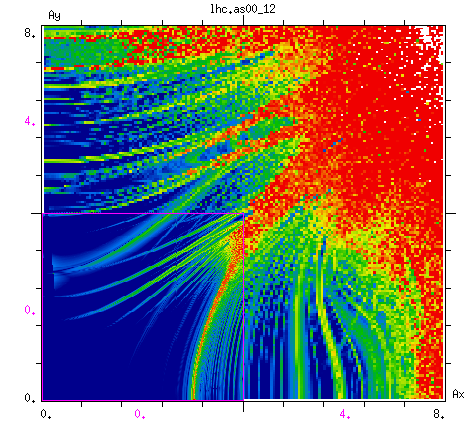}
	\end{minipage}
	\begin{minipage}[t]{0.3\linewidth}
		\centering
		10th turn pulsing
		\includegraphics[width=1.0\linewidth]{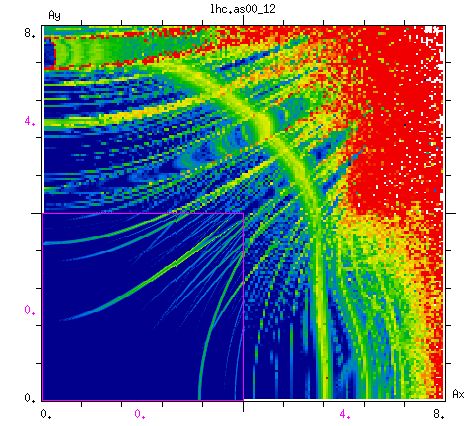}
	\end{minipage}
	\begin{minipage}[t]{0.3\linewidth}
		\centering
		2nd turn pulsing
		\includegraphics[width=1.0\linewidth]{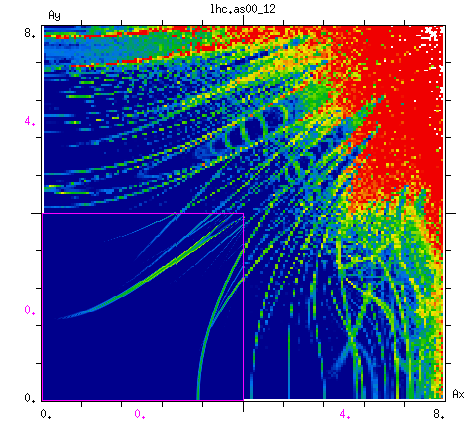}
	\end{minipage}
	\begin{minipage}[t]{0.3\linewidth}
		\centering
		3rd turn pulsing
		\includegraphics[width=1.0\linewidth]{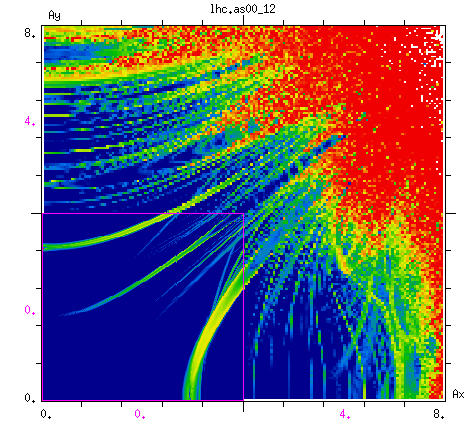}
	\end{minipage}
	\begin{minipage}[t]{0.3\linewidth}
		\centering
		4th turn pulsing
		\includegraphics[width=1.0\linewidth]{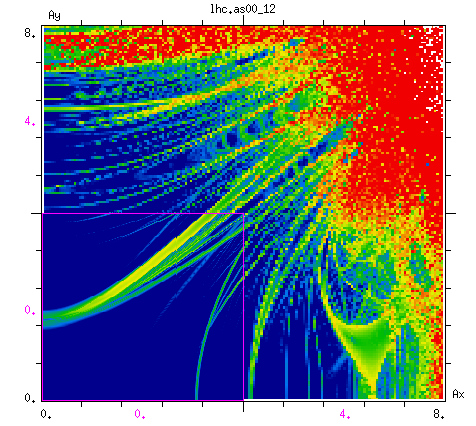}
	\end{minipage}
	\begin{minipage}[t]{0.3\linewidth}
		\centering
		5th turn pulsing
		\includegraphics[width=1.0\linewidth]{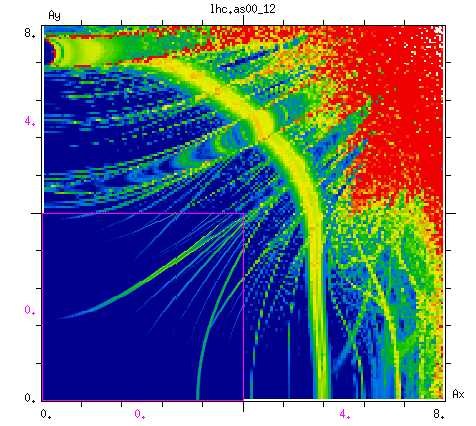}
	\end{minipage}
	\begin{minipage}[t]{0.3\linewidth}
		\centering
		6th turn pulsing
		\includegraphics[width=1.0\linewidth]{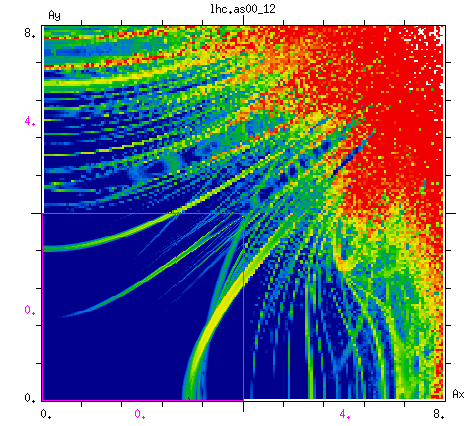}
	\end{minipage}
	\begin{minipage}[t]{0.3\linewidth}
		\centering
		8th turn pulsing
		\includegraphics[width=1.0\linewidth]{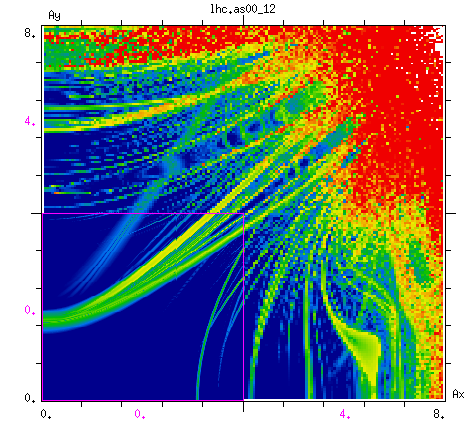}
	\end{minipage}
	\begin{minipage}[t]{0.3\linewidth}
		\centering
		9th turn pulsing
		\includegraphics[width=1.0\linewidth]{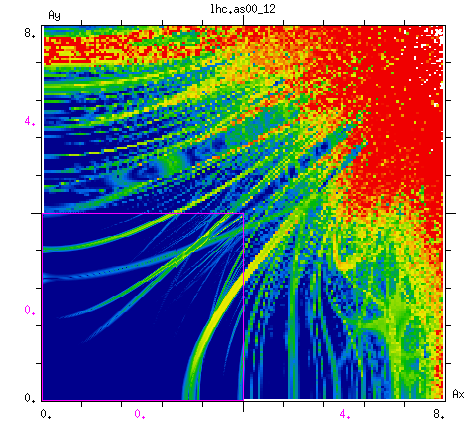}
	\end{minipage}
	\caption{Scenario 2 (with magnetic imperfections): FMA analysis in amplitude space for on-momentum particles ($\frac{\Delta p}{p_0}=0$) up to $8~\sigma$ amplitude for a square grid and 120~nrad excitation amplitude.\label{sim:fig:2:1:8}}
\end{figure}
\begin{figure}[h]
	\begin{minipage}[t]{0.3\linewidth}
		\centering
		no excitation
		\includegraphics[width=1.0\linewidth]{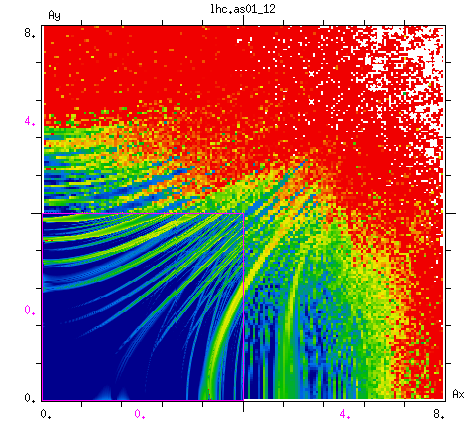}
	\end{minipage}
	\begin{minipage}[t]{0.3\linewidth}
		\centering
		7th turn pulsing
		\includegraphics[width=1.0\linewidth]{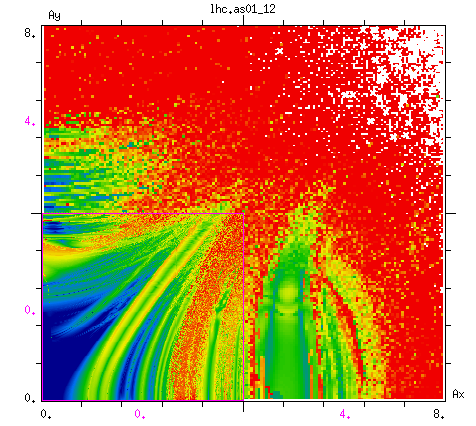}
	\end{minipage}
	\begin{minipage}[t]{0.3\linewidth}
		\centering
		10th turn pulsing
		\includegraphics[width=1.0\linewidth]{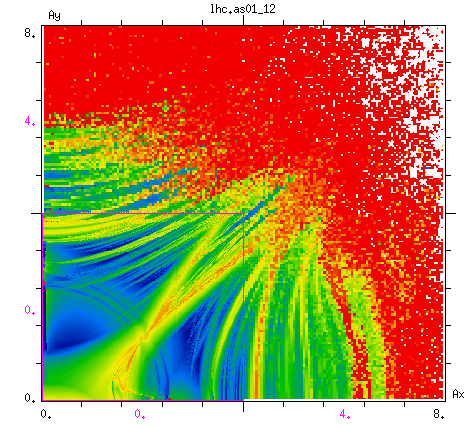}
	\end{minipage}
	\begin{minipage}[t]{0.3\linewidth}
		\centering
		2nd turn pulsing
		\includegraphics[width=1.0\linewidth]{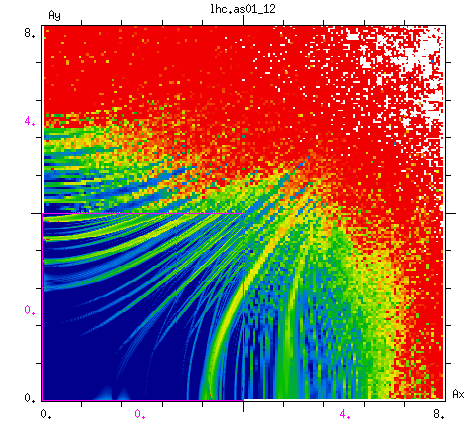}
	\end{minipage}
	\begin{minipage}[t]{0.3\linewidth}
		\centering
		3rd turn pulsing
		\includegraphics[width=1.0\linewidth]{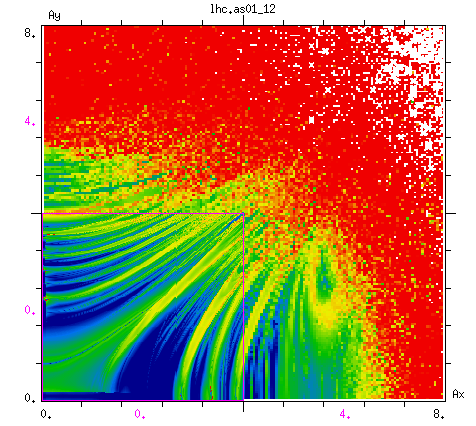}
	\end{minipage}
	\begin{minipage}[t]{0.3\linewidth}
		\centering
		4th turn pulsing
		\includegraphics[width=1.0\linewidth]{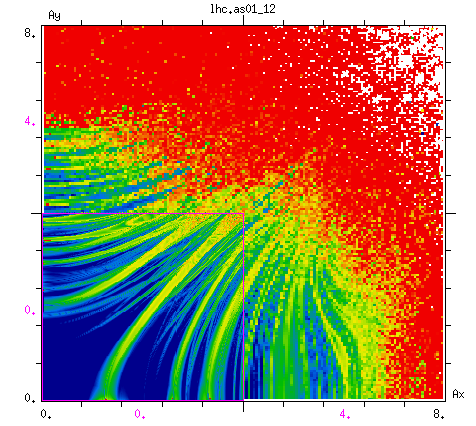}
	\end{minipage}
	\begin{minipage}[t]{0.3\linewidth}
		\centering
		5th turn pulsing
		\includegraphics[width=1.0\linewidth]{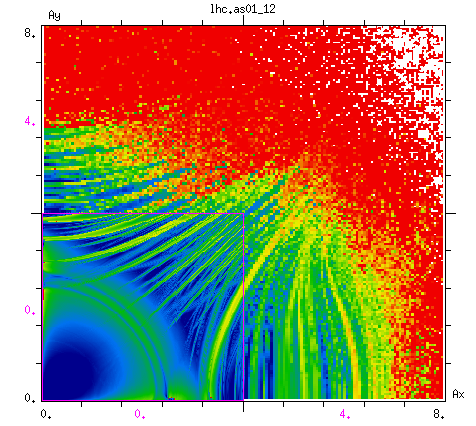}
	\end{minipage}
	\begin{minipage}[t]{0.3\linewidth}
		\centering
		6th turn pulsing
		\includegraphics[width=1.0\linewidth]{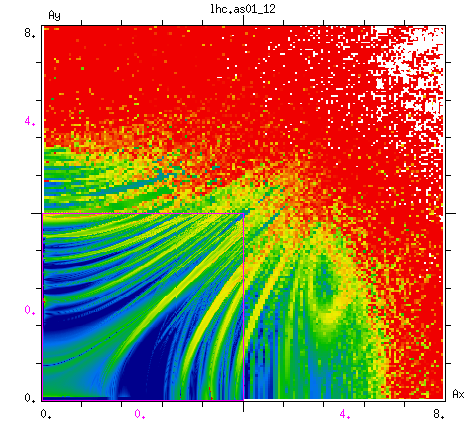}
	\end{minipage}
	\begin{minipage}[t]{0.3\linewidth}
		\centering
		8th turn pulsing
		\includegraphics[width=1.0\linewidth]{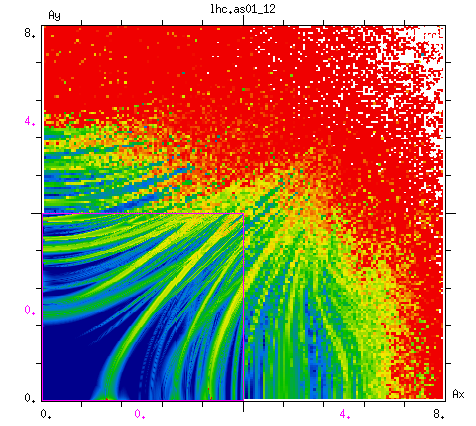}
	\end{minipage}
	\begin{minipage}[t]{0.3\linewidth}
		\centering
		9th turn pulsing
		\includegraphics[width=1.0\linewidth]{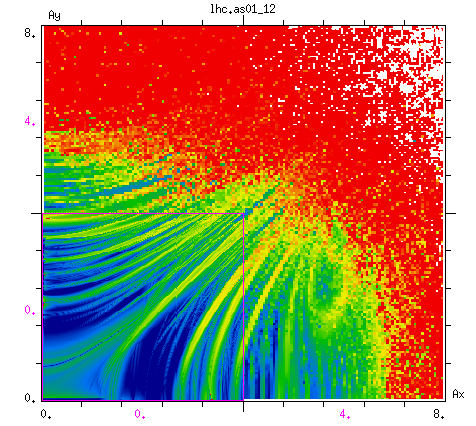}
	\end{minipage}
	\caption{Scenario 2 (with magnetic imperfections): FMA analysis in amplitude space for off-momentum particles ($\frac{\Delta p}{p_0}=0.1 \sigma_p$) up to $8~\sigma$ amplitude for a square grid and 120~nrad excitation amplitude.\label{sim:fig:2:1:9}}
\end{figure}

\clearpage
\section{Scenario 2 (with magnetic imperfections): Beam parameters for different pulsing patterns and excitation amplitude of 12~nrad}
\label{appendix:3}
\begin{figure}[h]
	\begin{minipage}[t]{0.49\linewidth}
		\centering
		\includegraphics[width=1.0\linewidth]{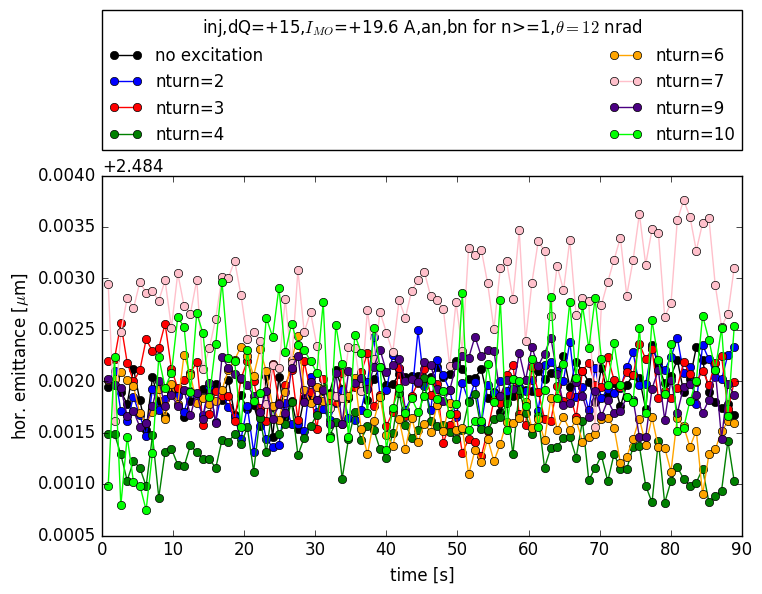}
	\end{minipage}
	\begin{minipage}[t]{0.49\linewidth}
		\centering
		\includegraphics[width=1.0\linewidth]{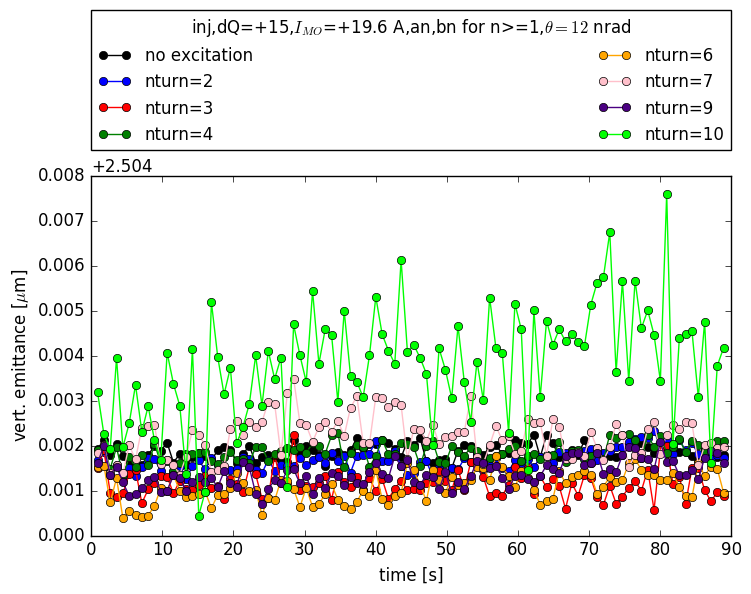}
	\end{minipage}
	\caption{Scenario 2 (with machine imperfections), excitation amplitude 12~nrad: Hor. (left) and Vert. (right) normalized emittance over $10^6$ turns. Emittance variations are negligibly small.\label{sim:fig:2:2:3}}
	\begin{minipage}[t]{0.49\linewidth}
		\centering
		\includegraphics[width=1.0\linewidth]{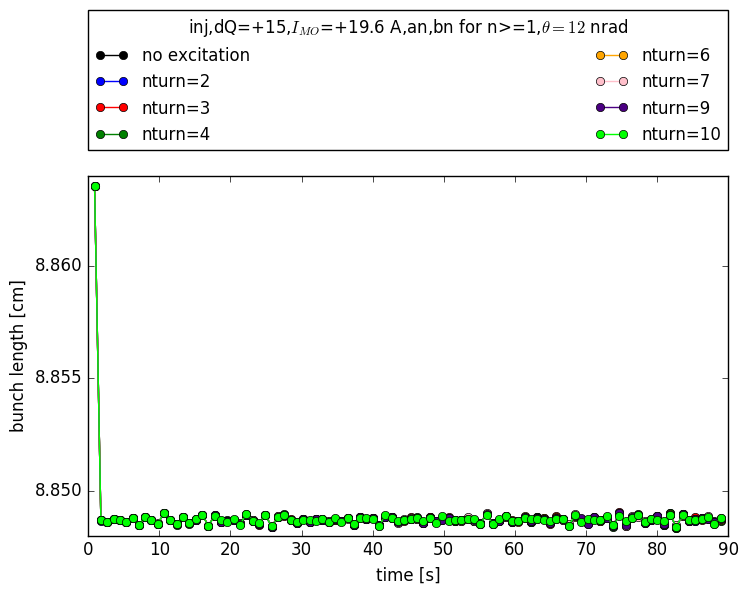}
	\end{minipage}
	\begin{minipage}[t]{0.49\linewidth}
		\centering
		\includegraphics[width=1.0\linewidth]{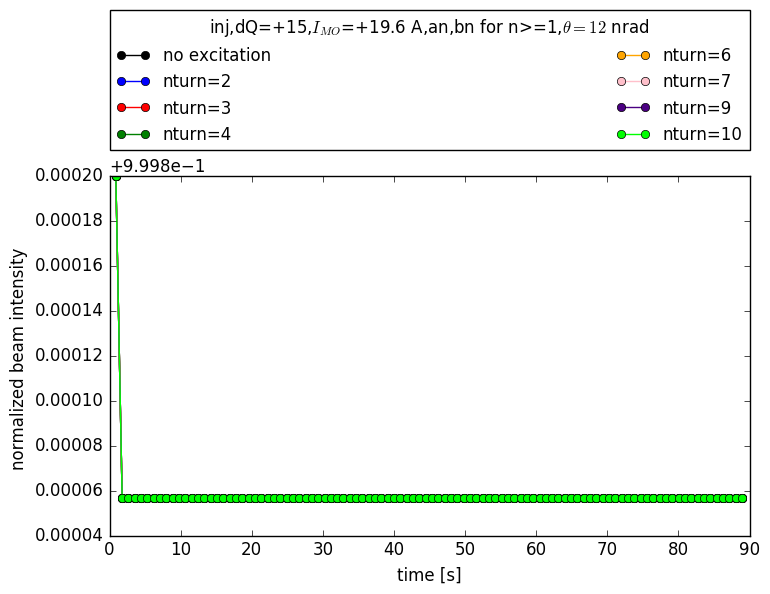}
	\end{minipage}
	\caption{Scenario 2 (with machine imperfections), excitation amplitude 12~nrad: $1\sigma$ Bunch length (left) and normalized beam intensity (right) over $10^6$ turns. No effects are observed at these excitation amplitudes.\label{sim:fig:2:2:4}}
\end{figure}






\clearpage

\bibliography{bibliography}

\end{document}